\documentclass[aps,prd,twocolumn,nofootinbib,floatfix,superscriptaddress]{revtex4-1}
\usepackage{setspace}
\usepackage{graphicx}
\usepackage{dcolumn}
\usepackage{multirow}
\usepackage{stackrel}
\usepackage{subfigure}
\usepackage{times,mathptm}
\usepackage{float}
\usepackage{color}
\usepackage{amsmath,amsfonts}
\usepackage{mathptmx}
\usepackage{mathrsfs}
\usepackage{bbm}
\usepackage{bm}
\usepackage{xfrac}

\newcommand{\beq}{\begin{equation}}
\newcommand{\eeq}{\end{equation}} 
\newcommand{\bea}{\begin{eqnarray}}
\newcommand{\eea}{\end{eqnarray}}

\def\lsim{\mathrel{\rlap{\lower4pt\hbox{\hskip1pt$\sim$}}
    \raise1pt\hbox{$<$}}}
\def\gsim{\mathrel{\rlap{\lower4pt\hbox{\hskip1pt$\sim$}}
    \raise1pt\hbox{$>$}}}
\newcommand{\D}{\mathcal{D}}

\newcommand{\J}{{\cal J}}

\renewcommand{\d}{\delta}

\renewcommand{\l}{\lambda}

\newcommand{\CD}{{\cal D}}

\newcommand{\tK}{\widetilde{K}}

\renewcommand{\b}{\beta}
\renewcommand{\a}{\alpha}

\renewcommand{\ni}{\noindent}
\newcommand{\tr}{\text{Tr}}

\renewcommand{\o}{\omega}

\newcommand{\vx}{{\vec{x}}}
\newcommand{\vy}{{\vec{y}}}
\newcommand{\vz}{\vec{z}}
\newcommand{\vu}{\vec{u}}
\newcommand{\vv}{\vec{v}}
\newcommand{\vw}{\vec{w}}
\newcommand{\vk}{{\vec{k}}}

\newcommand{\vR}{\vec{R}}

\newcommand{\q}{\overline{q}}

\renewcommand{\r}{\rho}
\newcommand{\e}{\epsilon}
\newcommand{\s}{\sigma}

\newcommand{\tD}{\widetilde{D}}

\newcommand{\M}{{\cal M}}

\newcommand{\oh}{\frac{1}{2}}

\newcommand{\dg}{\dagger}
\newcommand{\non}{\nonumber}

\newcommand{\rf}[1]{(\ref{#1})}
\newcommand{\ra}{\rightarrow}
\newcommand{\pa}{\partial}
\renewcommand{\vec}[1]{\bm #1}

\usepackage{ulem}

\begin{document}

\title{Constituent gluons and the static quark potential} 

\bigskip
\bigskip

\author{Jeff Greensite}
\affiliation{Physics and Astronomy Department, San Francisco State
University,   \\ San Francisco, CA~94132, USA}
\bigskip
\author{Adam P. Szczepaniak}
\affiliation{Department of Physics, Indiana University, Bloomington, IN 47405, USA}
\affiliation{Center for Exploration of Energy and Matter, Indiana University, Bloomington, IN 47403, USA}
\affiliation{Theory Center, Thomas Jefferson National Accelerator Facility, 12000 Jefferson Avenue, Newport News, VA 23606, USA}
\date{\today}
\vspace{60pt}
\begin{abstract}

\singlespacing
 
   We suggest that Hamiltonian matrix elements between physical states in QCD might be approximated,
in Coulomb gauge, by ``lattice-improved'' tree diagrams; i.e.\ tree diagram contributions with dressed ghost, transverse gluon, and Coulomb propagators obtained from lattice simulations.  Such matrix elements can be applied to a truncated-basis treatment of hadronic states which include constituent gluons.  As an illustration, we apply this hybrid approach to the heavy quark potential, for quark-antiquark separations up to 2.4 fm.  The Coulomb string tension in SU(3) gauge theory is about a factor of four times greater than the asymptotic string tension.  In our approach we show that a single constituent gluon is in principle sufficient, up to 2.4 fm, to reduce this overshoot by the factor required.  The static potential remains linear, although the precise value of the string tension depends on details of the Couloumb gauge ghost and gluon propagators in the infrared regime.  In this connection we present new lattice results for the transverse gluon propagator in position space. 

\end{abstract}

\maketitle

\singlespacing
\section{\label{intro}Introduction}

     It is not obvious that particle-like gluons, which are vital to perturbative QCD, really make sense as constituents of hadrons, particularly highly excited hadrons with higher spin.  Perhaps such states can only be described by strings of some kind which connect to quarks.  The picture of linear Regge trajectories as arising from a spinning line-like object is certainly compelling, and, if there is any sense in which gluons are constituents of hadrons, then surely the first challenge is to find out whether this line-like object has a substructure that can be understood in terms of individual gluons.  The simplest case to study is the lowest energy state containing a static quark-antiquark pair separated by a distance $R$. The line-like object connecting the quarks must manifest itself as a color electric flux tube. Does this flux tube have a substructure that involves individual gluons, as in the ``gluon chain'' proposal of ref.\ \cite{Greensite:2001nx}?

    To address this question, and in fact to even define what is meant by an individual gluon, it is necessary to work in a fixed gauge.  We will use Coulomb gauge, which has the advantage that a confining potential is already built into the 
dressed Coulomb propagator.  This fact has been verified numerically in many lattice studies 
\cite{Greensite:2014bua,Nakagawa:2006fk,Voigt:2008rr,Burgio:2012bk,Greensite:2004ke,Greensite:2003xf} and it is, moreover, a necessary condition for a non-vanishing asymptotic string tension \cite{Zwanziger:2002sh}.  The problem, however, is that the SU(3) Coulomb string tension $\s_{coul}$ derived from the instantaneous Coulomb propagator is a factor of four times larger than the asymptotic string tension \cite{Greensite:2014bua,Nakagawa:2006fk}, which seems too much of a good thing.  We may ask whether constituent gluons in a static quark-antiquark state can somehow reduce the Coulomb string tension to the known asymptotic value.   If that turns out to be true then we could go on to study other hadronic states, such as the low-lying glueballs, which would be another natural setting in which to investigate gluons as constituent particles.

    Among the physical states in Coulomb gauge, containing e.g.\  a static quark and antiquark pair, are superpositions of states of the form
\bea
   |\Psi \rangle_{\q q} &=& \int \prod_{i=1}^n d^3x_i ~\Psi_{k_1\ldots k_n}(x_1,x_2,\ldots,x_n)
\non \\
& &  \q^{\dag}(0) A_{k_1}(x_1) A_{k_2}(x_2)
         \ldots A_{k_n}(x_n) q^{\dag}(R) |0 \rangle_{\text{vac}} \ ,
\label{chain}
\eea
where $ |0 \rangle_{\text{vac}}$ is the true vacuum state, $q^{\dag},\q^{\dag}$ are massive quark-antiquark creation operators, and $\Psi$ is a function which, in a variational approach, may depend on some set of parameters. Suppose we have a finite, not necessarily orthogonal, set of such states, labeled by an integer $\{|j \rangle \}$.  From these a set 
$\{|\tilde{j} \rangle \}$ of orthogonal states can be constructed.  If we could compute Hamiltonian matrix elements 
$\langle \tilde{j} | H | \tilde{k} \rangle$ in the Hilbert space spanned by this truncated basis, then the standard procedure is to diagonalize the Hamiltonian in the truncated basis, minimize the energy of the lowest energy state by adjusting the variational parameters, and in this way arrive at an estimate for the static quark potential at separation $R$.  A similar strategy could be employed in spectrum calculations, involving states with dynamical quarks. The problem, of course, is to calculate the relevant overlaps and Hamiltonian matrix elements.  In principle this task can be carried out by lattice Monte Carlo in Coulomb gauge, and this was the path followed in \cite{Greensite:2009mi}.  The problem was also addressed from the Dyson-Schwinger point of view in 
\cite{Szczepaniak:2005xi}.  In this article we will suggest a somewhat different approach, inspired by renormalized perturbation theory.

    Consider the expression
\beq
           C_{jk}(t) = \langle j | e^{-Ht} |k \rangle  \ .
\eeq
The time derivative evaluated at $t=0$ gives us $ \langle j | H |k \rangle$, while $C_{jk}(0)$ is
the overlap $\langle j |k \rangle$.  The prescription for calculating $C_{jk}(t)$, in ordinary perturbation theory, is essentially the same as the prescription for calculating an S-matrix element:  Sum all of the tree diagrams which contribute to this expression, including all $n$-point vertices.  The vertices are the one-particle irreducible (1PI) n-point functions appearing in the quantum effective action, and the propagators are the full (or ``dressed'') propagators of the theory.  The task of perturbation theory is to compute these renormalized propagators and 1PI n-point functions. This program can be carried out analytically, within the limits of validity of an asymptotic expansion, if the spatial separations of the quarks and gluons in states $\{|j \rangle\}$ are small.  If this is not the case then the program fails,  because the perturbative expansion for the relevant propagators and vertices rapidly diverges.  
Let us observe, however, that only a finite number of tree diagrams contribute to the calculation of $C_{jk}(t)$. The sum of trees is not an infinite expansion; it is finite and, given the dressed propagators and n-point vertices, it supplies the exact answer, regardless of the magnitude of the renormalized coupling. Therefore, if propagators and relevant vertices could be calculated by some non-perturbative approach, say by Dyson-Schwinger equations or lattice Monte Carlo simulations, then the tree diagrams could be summed, and $C_{jk}(t)$ could be calculated.  From those quantities, the spectrum of the
Hamiltonian in the subspace of Hilbert space spanned by the set $\{|j \rangle\}$ could be  calculated.

    In this article we will take a first step along these lines, by taking ghost, transverse gluon, and Coulomb propagators from lattice Monte Carlo simulations, neglecting all vertices apart from those arising from the non-polynomial operator in the Coulomb gauge Hamiltonian.  We will use the resulting tree diagrams to compute Hamiltonian matrix elements 
in a truncated basis of thirteen states, consisting of one state with no constituent gluons, and twelve states with a single constituent gluon in various spatial distributions.  It is found that the static quark potential derived in this way remains linear, but the asymptotic string tension depends on an overall constant factor associated with the
ghost propagator.  In the absence of decisive data on this point, we simply tune the factor in the ghost propagator to get the known result.  Hopefully future lattice Monte Carlo investigations of the ghost propagator will make our study more
predictive, but for now we only show that inclusion of constituent gluons in static quark-antiquark states, along the lines of the gluon chain model, provides a very plausible mechanism for reducing the string tension from the pure Coulombic
value, which is much too high, to the value consistent with numerical simulations.

    Below in Section \ref{review} we review some of the numerical results and conjectures in ref.\ \cite{Greensite:2014bua},
which motivate the work presented here.  In Section \ref{LIT} we will present our proposal in detail, and in particular explain how the non-local operator which appears in the Coulomb gauge Hamiltonian is treated in the tree diagram approach.  Expressions for the Hamiltonian matrix elements in terms of lattice-improved tree diagrams are derived in section \ref{matrix}, and they require, in addition to the
Coulomb propagator already obtained in \cite{Greensite:2014bua}, also the ghost and transverse gluon propagators. In Section  \ref{gluons} we will show our lattice Monte Carlo results for the equal times transverse gluon propagator in position space, relying on refs.\ \cite{Burgio:2008jr,Burgio:2012bk,Nakagawa:2009zf} for the infrared behavior of
the ghost propagator.  In Section \ref{st} we will bring all these results together, compute an estimate for the static quark potential, and show how the superposition of zero and one constituent gluon states can bring the static quark potential down from the Coulomb result by a large numerical factor dependent on the ghost propagator.  We conclude in Section \ref{conclude}.  Finally, in the Appendix we discuss the results of Monte Carlo simulations of gluon, ghost and Coulomb propagators in the context of Dyson-Schwinger approach.

\section{\label{review}Coulomb potential and the gluon chain model}

    In a recent article \cite{Greensite:2014bua} we calculated the non-perturbative Coulomb potential in SU(3) pure gauge theory via lattice Monte Carlo simulations in Coulomb gauge, and found it to be
\beq
           V_{coul} (R) =  \s_{coul} R -  {\pi \over 12}{1\over R} \ ,
\label{cpot}
\eeq
where the Coulomb string tension is 
\bea
              \s_{coul} &\approx& 20.5(4) ~ \text{fm}^{-2}
\non \\
                      &\approx& (893\pm 9~\text{MeV})^2 \ ,
\label{cpot2}
\eea
(see also \cite{Nakagawa:2006fk}) which is about four times the accepted value of the asymptotic string tension $\s = (440~\text{MeV})^2$.  This is clearly too much of a good thing.  While it is helpful that a linearly confining potential is obtained in Coulomb gauge by what amounts to one (dressed) gluon exchange, it is not acceptable that the static quark potential is four times too large.  Moreover there is no indication that the color electric field generated by the Coulomb propagator around the quark-antiquark sources is restricted to a flux tube.  Something else important must be going on.

    The key point is that $V_{coul}(R)$ is not necessarily the minimal energy of a state containing a pair of static
quark-antiquark sources.  It is, rather, the interaction energy of a specific state in Coulomb gauge, namely
\beq
           |0\rangle_{\q q} = \q^{\dag}(0) q^{\dag} (R) |0\rangle_{\text{vac}} \ ,
\label{0g}
\eeq 
where 
\beq
          \Phi_0[A] = \langle A |0\rangle_{\text{vac}}
\eeq
is the true vacuum wave functional.  The energy of this state, including self-energies, is given by the logarithmic time derivative of the
Euclidean-time correlator
\beq
       V(R) = - \lim_{t\ra 0} {d \over dt} \log {}_{\q q}\langle 0| e^{-Ht} ||0\rangle_{\q q} \ .
\eeq
 As Dirac indices and quark kinetic energies are not relevant to our study, it is sufficient to compute, in a Euclidean action formulation, the logarithmic time derivative of a correlator of short timelike Wilson lines
\beq
           V(R) = - \lim_{t\ra 0} {d \over dt} \log  \big\langle \tr[L_t({\bf 0}) L_t^\dg({\bf R})] \big\rangle \ ,
\eeq
where
\beq
              L_t(\vx) \equiv T\exp\left[ig\int_0^t dt A_0(\vx,t) \right] \ .
\eeq
In a lattice formulation, the calculation of $V(R)$ boils down to to calculating the logarithm of the vacuum expectation 
value (VEV) of products of two timelike links, evaluated at equal times.  This method for computing the Coulomb energy was first suggested in \cite{Greensite:2003xf}, and the potential defined in this way includes an $R$-independent self-energy term.  The self-energy term, proportional to the inverse lattice spacing, can be identified and subtracted away, with the result for the $R$-dependent quark-antiquark interaction potential $V_{coul}(R)$ shown above
in \rf{cpot} and \rf{cpot2}.  The coefficient $\pi/12$ in \rf{cpot} is the correct value for the L\"uscher term, but this coefficient is not simply assumed.  Rather, it appears to be the likely continuum limit of the values derived at finite lattice spacings, cf.\  \cite{Greensite:2014bua}.

\begin{figure*}[t]
\subfigure[]  
{   
 \label{kinetic}
 \includegraphics[scale=0.32]{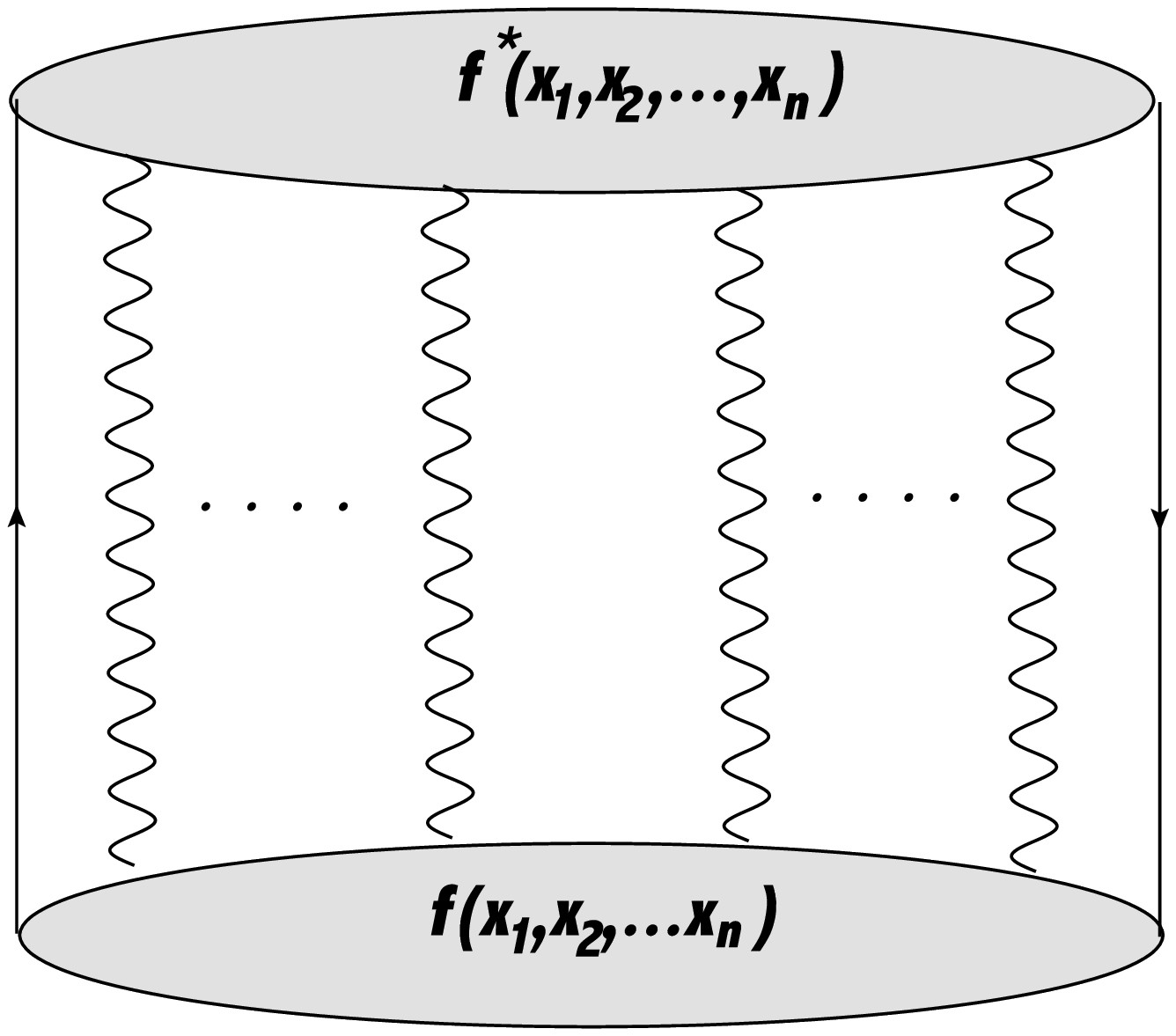}
}
\subfigure[]  
{   
 \label{coulomb}
 \includegraphics[scale=0.32]{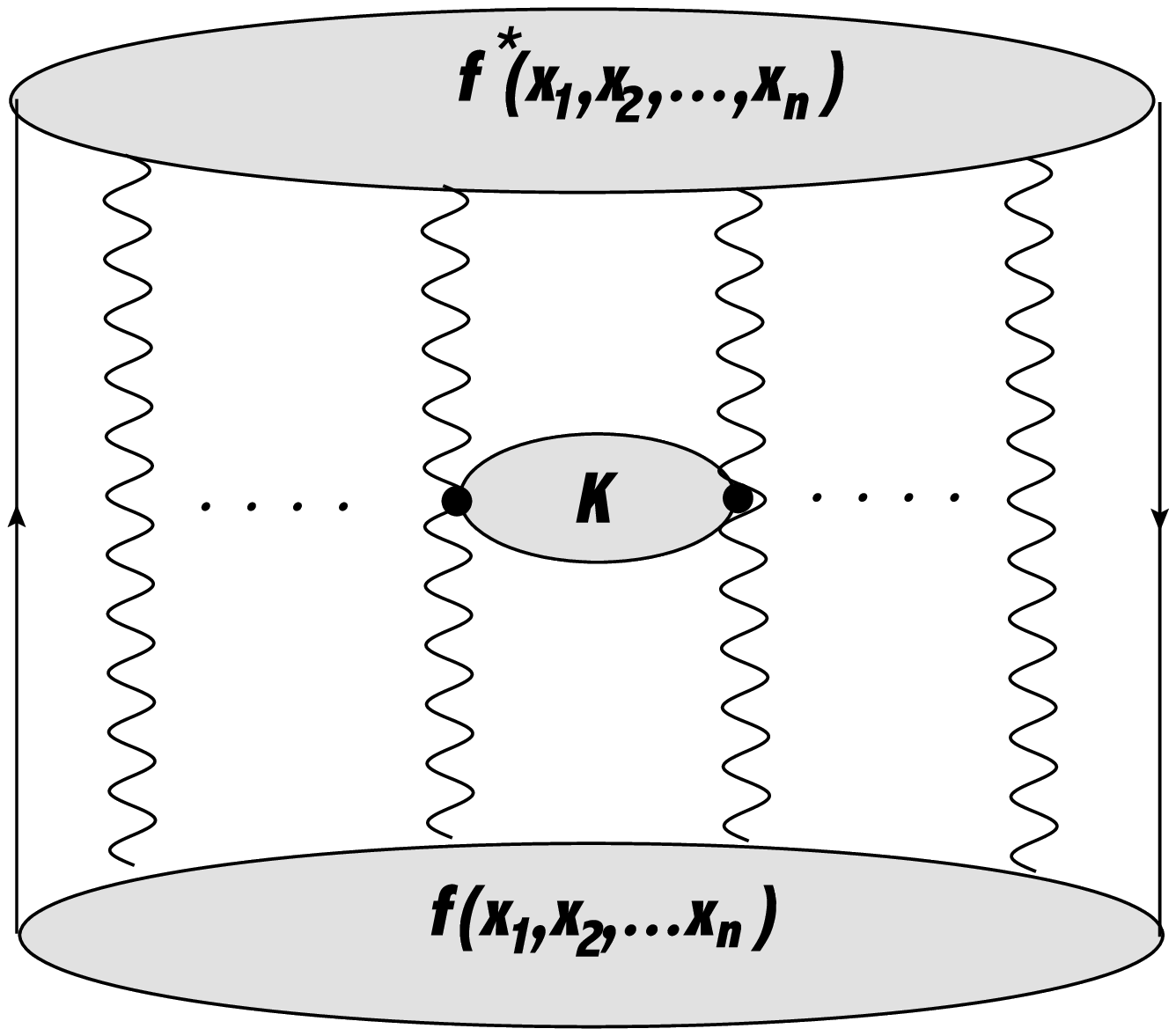}
}
\subfigure[]  
{   
 \label{offdiagonal}
 \includegraphics[scale=0.32]{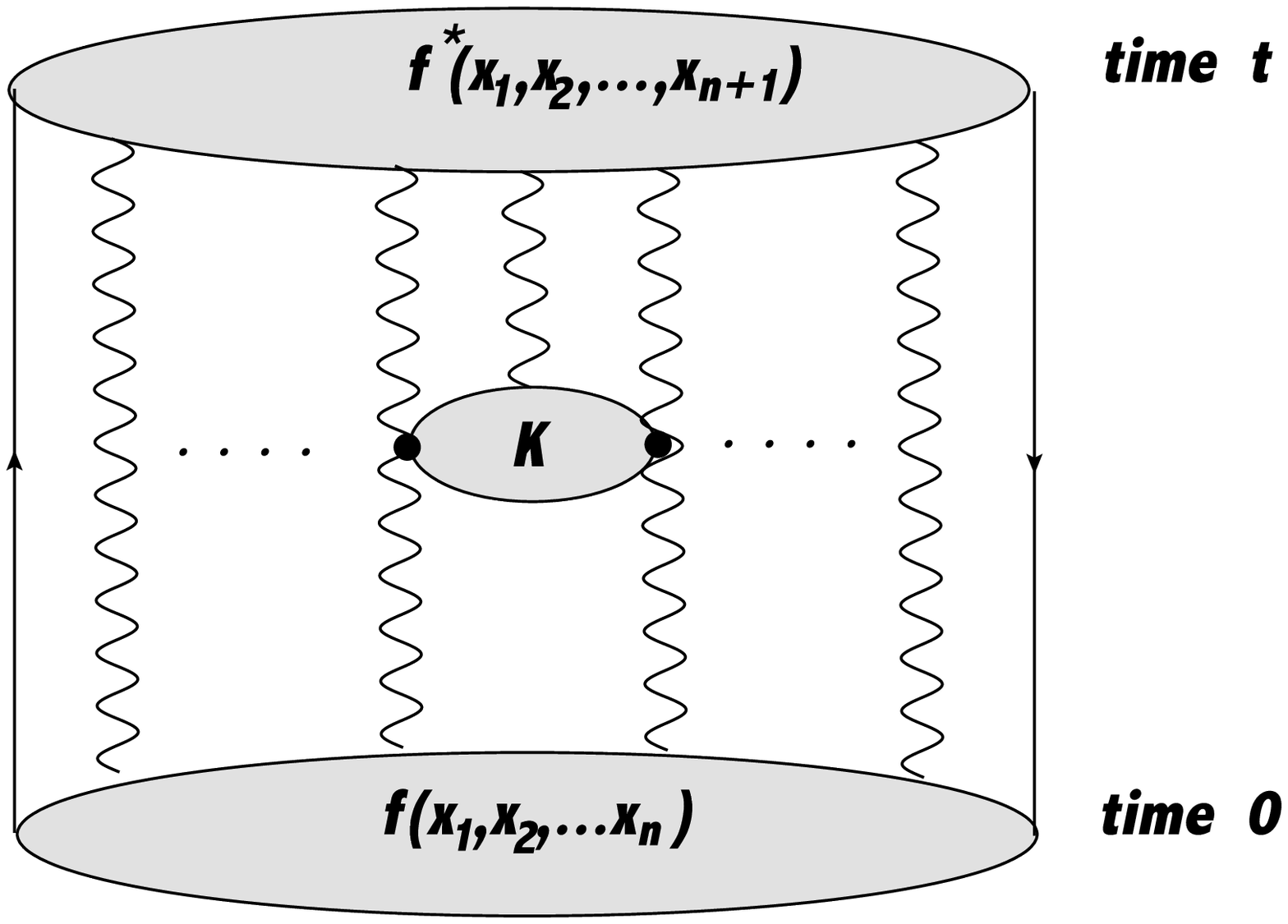}
}
\caption{Diagrams which, after a time derivative, contribute to Hamiltonian matrix elements.  (a) The graph which determines the kinetic energy of constituent gluons. (b) One of the graphs determining the Coulomb energy of an 
$n$-gluon state.  The blob labeled ``K'' is the instantaneous non-polynomial Coulomb operator.  (c) Schematic of a graph contributing to an off-diagonal Hamiltonian matrix element between states with $n$ and $n+1$ gluons.  
Here one of the $A$-field operators in the Coulomb operator $K(A)$ contracts with a gluon in the final state.}
\label{graphs}
\end{figure*}

    Since the state $|0\rangle_{\q q}$ cannot be the minimal energy state containing static quarks, it is reasonable
to consider states with $n$ constituent gluons of the form shown in \rf{chain}, where what we mean by $n$
``constituent gluons'' is simply that there are $n$ $A$-field operators that operate on the true ground state.  In diagramatic
representation this ket vector corresponds to $n$ transverse gluon lines emerging from a blob at time $t_0=0$, and the
bra vector is a blob with $n$ transverse gluons lines entering at some later time $t$.  The static quark and antiquark lines
attach to either end of the blob.  The gluon chain model of ref.\ \cite{Greensite:2001nx} proposed that the minimal
energy state containing a static quark antiquark pair consists of some number $n$ of constituent gluons arranged
roughly in a line (or cylindrical region) between the quark and antiquark, with the color ordering of the gluons correlated with the spatial ordering of the gluons along the line.  The original motivation was the observation that the force between
colored sources, in two-loop perturbation theory, grows very rapidly with $R$ when the running coupling approaches values of O(1).  So the idea is that as a quark and antiquark separate, the energy grows until at some point it is energetically favorable to place a gluon in between the quark and antiquark, which halves the effective color charge
separation.  As the quark and antiquark continue to separate, eventually it is energetically favorable to introduce a
second gluon between the two sources, and so on.   In the end, the minimal energy state would contain
approximately $n=R/R_0$ constituent gluons, where $R_0$ is essentially the average distance between gluons, and
the kinetic plus inter-gluon interaction energy is $E_0$.  Then the total energy of the long chain is approximately
$E(R) = n E_0 = (E_0/R_0) R$, which is a linear potential with string tension $E_0/R_0$.

    It now seems clear that this picture is untenable, because it assumes that at some point, as color charges separate,
the potential between colored sources grows faster than linear.  But if the Coulomb potential between gluons is instead asymptotically linear, then there is no advantage to introducing more gluons.  If the gluons were arranged exactly in a line
between the quarks, and the inter-gluon Coulomb potential is linear with string tension $\s'$, then the 
overall Coulomb energy would be $\s' R$ regardless of the number of gluons in the chain.  The kinetic energy of the
gluons, and the transverse fluctuations away from the line joining the quark-antiquark pair, could only increase this
energy.  So it would appear that the minimal energy really is the zero constituent gluon state \rf{0g}, and we have found
that this state has a string tension which is far above the asymptotic string tension.

   However, this conclusion ignores the fact that a state with a fixed number of constituent gluons is not an eigenstate
of the Hamiltonian, and there will always be non-zero matrix elements $\langle m|H|n \rangle$ between states with
different numbers of constituent gluons.  This means that the lowest energy state is certain to be a superposition of states with different numbers of constituent gluons.  Diagonal matrix elements contain both purely kinetic contributions,
indicated schematically in Fig.\ \ref{kinetic}, and instantaneous Coulomb interactions, shown in Fig.\ \ref{coulomb}.  Matrix elements between states with different numbers of constituent gluons are associated with diagrams such as Fig.\ \ref{offdiagonal}. All matrix elements are derived from the time derivatives of the diagrams indicated.  In ref.\ \cite{Greensite:2014bua} we showed that if we make some plausible assumptions about the behavior of the diagonal and off-diagonal Hamiltonian matrix elements, then the static quark potential associated with the lowest energy state could be reduced from the purely Coulomb value by a large numerical factor, while retaining the asymptotic linearity of the potential.  For details of the model calculation we refer the reader to \cite{Greensite:2014bua}.  In the present article we will actually evaluate the type of diagrams just indicated, in a ``lattice-improved''  tree-diagram framework, to see if they really do have the conjectured effect.

\section{\label{LIT}Lattice improved tree diagrams for the static quark potential}

\subsection{Preliminaries}

   For completeness and to establish notation we begin with the usual preliminaries regarding Coulomb gauge.
The Coulomb gauge Hamiltonian is $H=H_{glue}+H_{coul}+H_{matter}$, where
\begin{eqnarray}
        H_{glue} = \oh \int d^3x\;( \J^{-\oh}\vec{E}^{{\rm tr},a} {\cal J}
        \cdot \vec{E}^{{\rm tr},a} \J^{-\oh} + \vec{B}^a \cdot \vec{B}^a),
\non \\
        H_{coul} = \oh \int d^3x d^3y\;\J^{-\oh}\rho^a(x) \J
                K^{ab}(x,y;A) \rho^b(y) \J^{-\oh},
\non \\
\end{eqnarray}
with 
\begin{eqnarray}
        K^{ab}(x,y;A) &=& \left[ \M^{-1}
        (-\nabla^2) \M^{-1} \right]^{ab}_{xy},
\non \\
        \rho^a &=& \rho_q^a  + \rho^a_{\bar q}   + \rho^a_g  ,
\non \\
         \M &=& -\nabla \cdot \CD(A) ~~,~~ \J = \det[\M].
\end{eqnarray}
Here  $\rho^a_q(x) = g  q^\dag_i(x) t^a_{ij} q_j(x)$, $\rho^a_{\bar q}(x) = g  \bar q_i(x) t^a_{ij} \bar q^{\dag}_j(x)$ and  
$\rho^a_g(x) = - g f^{abc} A^b_k(x) E^c_k(x)$
are the  charge density of quarks, antiquarks and gluons, respectively, and $\CD_k(A)$ is the covariant derivative.
$H_{matter}$ is the part of the Hamiltonian containing dynamical matter fields.  It will not be needed here, since
in this article we are only concerned with static color sources, which can be represented by Wilson lines in the time direction.  The operator-ordering terms $\J^{\pm \oh}$ do not appear at the classical level, and therefore do not appear in the construction of tree diagrams.  

   An equivalent Euclidean path-integral formulation in first-order formalism \cite{Zwanziger:1998ez,Feinberg:1977rc,Cucchieri:2000hv,Watson:2007mz} can be based on the generating functional
\bea
\lefteqn{Z[J] =}
\non \\
& & \int_G DA_i^{tr} \int DE_i^{tr} \exp\left[ \int d^4x \Bigl(iE_i \dot{A}_i -  \oh (E_i^{2}  + B_i^2) 
-  i  J_i A_i \Bigr) \right.
\non \\
& & \left. - \oh \int dt d^3x d^3y (\rho_g + gJ_4)_{\vx,t} K[\vx,\vy,t,A]  (\rho_g +  J_4)_{\vy,t} \right] \ ,
\non \\
\eea
where the ``$tr$'' superscript in the measure indicates that the integration is restricted to transverse $A,E$ fields, and
$\rho^a_g(x) = -g f^{abc} A_i^{b}(x) E_i^{c}(x)$.

   We define the transverse gluon propagator
\bea
        D^{ab}_{ij}(x,t) &=& \langle A_i^a(x,t) A_j^b(0,0) \rangle 
\non \\
          &=& D_{ij}(x,t) \d^{ab} \ ,
\eea
the $E-A$ propagator
\bea
\tD_{ij}^{ab}(x,t) &=& \langle E_i^a(x,t) A_j^b(0,0) \rangle
\non \\
&=& \tD_{ij}(x,t) \d^{ab} \ ,
\eea
the ghost propagator
\bea
       G^{ab}(\vx-\vy) &=& \left\langle \left({1 \over -\nabla \cdot \CD}\right)_{xy}^{ab}\right\rangle
\non \\
&=& G(x-y) \d^{ab} \ ,
\eea
and the $K$-propagator
\bea
       K^{ab}(\vx-\vy) &=&  \left\langle \left({1 \over -\nabla \cdot \CD}\right)_{xz}^{ac}(-\pa^2)_z
           \left({1 \over -\nabla \cdot \CD}\right)_{zy}^{ac}\right\rangle
\non \\
      &=& \d^{ab} K(\vx-\vy)  \ .
\eea

   The Coulomb propagator demands some special attention; it is not the same as the $K$-propagator.  It is defined as
\bea
\lefteqn{D^{ab}_{44}(x-y)} & &
\non \\
 & & \qquad = \left({1\over Z} {\d^2 \over \d J_4^a(x) \d J_4^b(y)} Z \right)_{J=0}
\non \\
& & \qquad = \langle K^{ab}(\vx-\vy;A) \rangle \d(x_4-y_4)
\non \\
& & \qquad  + 
          \int d^3z d^3w \langle K^{ac}(\vx-\vz) \rho^c_g(z) \rho^d_g(w) K^{cb}(\vw-\vy) \rangle
\non \\
& & \qquad = \d^{ab}\bigg( K(\vx-\vy) \d(x_4-y_4) + P(x-y) \bigg) \ .
\eea
In $d=4$ dimensions $P(x-y)$ may have both an instantaneous and non-instantaneous part
\beq
P(x-y) = P_{inst}(\vx-\vy) \d(x_4-y_4) + P_{non}(x-y) \ ,
\eeq
so that
\bea
 D^{ab}_{44}(x-y) &=&
  \d^{ab}\bigg( K(\vx-\vy)   + P_{inst}(\vx-\vy) \bigg) \d(x_4-y_4)
\non \\
  & & + P_{non}(x-y) \ .
\eea
The instantaneous part of the Coulomb propagator will be denoted
\bea
\tK^{ab}(x-y) &=& \d^{ab} \tK(\vx-\vy)
\non \\
&=& \d^{ab} \bigg( K(\vx-\vy) + P_{inst}(\vx-\vy) \bigg) \ .
\eea
It was shown by Zwanziger \cite{Zwanziger:1998ez,Cucchieri:2000hv} that both $g^2 D^{ab}_{44}(x-y)$ 
and $g^2 \tK(\vx-\vy)$ are renormalization group
invariants.  For now only the instantaneous part of the Coulomb propagator will be needed.  We will denote the
relationship of $\tK$ to operator expectation values by
\beq
   \tK = \langle K \rangle + \langle K \rho \rho K \rangle_{inst} \ ,
\label{tK}
\eeq
where the subscript on the last term indicates the instantaneous part of the VEV.  The relationship between the
Coulomb potential between static sources in the fundamental representation and the instantaneous Coulomb 
propagator is
\beq
    V_{coul}(R) = - g^2 C_F \tK(R) \ .
\label{VK}
\eeq
For later use we will introduce the notation
\bea
           K_c(R) &=& V_{coul}(R)
\non \\
           G_c(R) &=& g G(R) \ .
\eea

\begin{figure}[t!]
\centerline{\scalebox{0.6}{\includegraphics{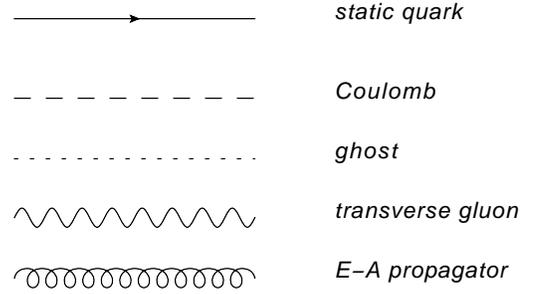}}}
\caption{Diagrammatic notation for Coulomb gauge propagators.}
\label{legend}
\end{figure}

Our graphical notation for these propagators is shown in Fig.\ \ref{legend}.  We will assume that the
transverse gluon propagator has the form
\bea
D_{ij}(x,t) &=& \int {d^4k \over (2\pi)^4} {e^{i(k_0 t + \vk \cdot \vx)} \over k_0^2 + \o_k^2} \left( \d_{ij} - {k_i k_j \over k^2} \right)
\non \\
&=& \int {d^3k \over (2\pi)^3} {e^{i \vk \cdot \vx - \o_k t} \over 2\o_k} \left( \d_{ij} - {k_i k_j \over k^2} \right)
\non \\
&=& \d_{ij} D(x,t) - D^{(1)}_{ij}(x,t)
\non \\
D(x,t) &=& \oh \d_{ij} D_{ij}(x,t) \ ,
\label{gluonprop}
\eea
(where $k^2 \equiv \vk \cdot \vk$), and also that
\bea
\tD_{ij}(x,t )&=&  \int {d^4k \over (2\pi)^4} {k_0 e^{i(k_0 t + \vk \cdot \vx)} \over k_0^2 + \o_k^2} \left( \d_{ij} - {k_i k_j \over k^2} \right)
\non \\
&=& \d_{ij} \tD(x,t) - \tD^{(1)}_{ij}(x,t)
\non \\
\tD(x,t) &=& \oh \d_{ij} D_{ij}(x,t) \ .
\eea
At the perturbative level $\o_k = |\vk|$.  At the non-perturbative level $\o_k$ will be something else, with   
\beq
          \o_k = \left\{ \begin{array}{cl}
                   \sqrt{k^2 + m^2} & \mbox{massive propagator} \\
                   \sqrt{k^2 + m^4/k^2} & \mbox{Gribov propagator} \end{array} \right.
\label{omega}
\eeq
as possible candidates.   Our approach is to determine the relevant gluon propagators in position space from lattice Monte Carlo simulations.  We will not need to know $\o_k$  explicitly, although we will, in Section \ref{gluons}, 
compare our numerical results for the transverse gluon propagator in position space with the forms implied by the above candidates for $\o_k$.   However, the assumption that $D_{ij},\tD_{ij}$ have the form above allows us to conclude that
\bea
D'(x,0) &\equiv& \lim_{t=0} {d \over dt} D(x,t)
\non \\
&=& -\oh \d^3(x)
\non \\
\lim_{\e \ra 0} \tD(x,\e) &=& {i \over 2} \text{sign}(\e) \d^3(x) \ .
\eea

\subsection{Decomposition of the Coulomb vertex}

    In the tree diagram framework we suggest here, there is a question of how the non-polynomial operator 
\beq
\rho^a(x) K^{ab}(x-y;A) \rho^b(y) 
\eeq 
\ni should be handled.  Let us first consider the case where the charge operators $\r$ contract only with ``external'' operators in the initial and final states, either heavy quarks or constituent gluons.  These operators
will be denoted  $\rho^{ext}$.  Then we also have to consider the possibility that some $A$-field operators in the perturbative expansion of $K^{ab}(x-y;A)$ contract with
gluon operators in the initial and final states.  In some diagrams there are no such contractions, and those diagrams 
sum up to the dressed $K$-propagator $\langle K \rangle$.  We must also consider the product of $\rho K \rho$ operators in which two charge operators contract with initial and final states, i.e. $\rho^{ext} K\rho \rho K \rho^{ext}$.  Again there are diagrams in which there are no contractions of operators in either $K$ with the external constituent gluons.  Those diagrams sum up to $\langle K \rho \rho K \rangle$, and add to the $K$-propagator $\langle K \rangle$ to produce 
the dressed Coulomb propagator $\tK(x-y)$ (see \rf{tK}).  

    In other diagrams, however, it is necessary to consider
a ``Coulomb vertex,'' as indicated schematically in Fig.\ \ref{offdiagonal}, where one or more field operators in $K$
contract with operators in the initial and/or final states.  It turns out, as we will now show, that Coulomb vertices can be decomposed into products of ghost operators, Coulomb operators, and transverse gluon operators that contract with the external states.  In our tree diagram formulation, these ghost operators and Coulomb operators just become dressed ghost and Coulomb propagators, and this prescription amounts to a partial resummation of the full perturbation series.

   We will begin with the case $\rho^{ext} K \rho^{ext}$, so it is sufficient to just consider the expansion of the Coulomb operator
\bea
K^{ab}_{\vx \vy}(A) &=& G^{ac}_{\vx \vz} (-\nabla^2) G^{cb}_{\vz \vy} \ ,
\eea
where the ghost operator is
\bea
G^{ab}_{\vx \vy} &=& (\M^{-1})^{ab}_{\vx \vy} =  \left({1\over -\nabla \cdot D}\right)_{\vx \vy}^{ab}
\non \\
(-\nabla \cdot D)^{ab} &=& -(\d^{ab} \nabla^2 + gf^{abc} A^b_i \pa_i) \ ,
\eea
The perturbative expansion of the ghost operator begins with 
\beq
         -\nabla^2 G^{ab}_{\vx \vy} = \d^{ab}\d^3(\vx-\vy) + gf^{acd}A^c_i(\vx) \pa_i G^{db}_{\vx \vy}
\eeq
or
\beq
           G^{ab}_{\vx \vy} = \d^{ab} \left({1 \over (-\nabla^2)}\right)_{\vx \vy} + \left({1 \over (-\nabla^2)}\right)_{\vx \vz}
                 gf^{acd}A^c_i(\vz) \pa_i G^{db}_{\vz \vy} \ .
\eeq
This equation can be solved iteratively, and from here on we will drop both color and spatial indices.  The solution is a power series
\bea
          G &=& {1 \over (-\nabla^2)} \sum_{n=0}^\infty \left(gfA_i\pa_i {1 \over (-\nabla^2)}\right)^n
\non \\
              &=& {1 \over (-\nabla^2)} \sum_{n=0}^\infty M^n \ ,
\eea
where we define
\beq
          M = gfA_i\pa_i {1 \over (-\nabla^2)} \ .
\eeq
The Coulomb operator is then
\bea
          K &=&  {1 \over (-\nabla^2)} \sum_{m=0}^\infty M^m (-\nabla)^2 {1 \over (-\nabla^2)} \sum_{n=0}^\infty M^n
\non \\
             &=& {1 \over (-\nabla^2)} \sum_{m=0}^\infty \sum_{n=0}^\infty M^{m+n}
\non \\
             &=& {1 \over (-\nabla^2)} \sum_{N=0}^\infty (N+1) M^N \ .
\eea
 
\begin{figure}[t]
\centerline{\scalebox{0.5}{\includegraphics{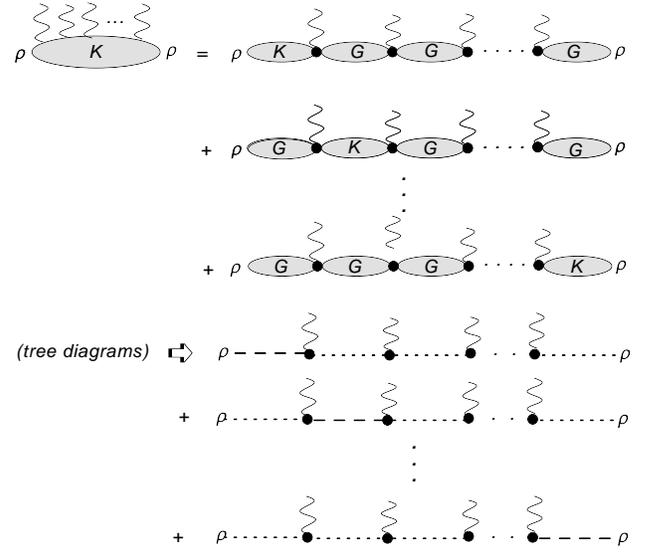}}}
\caption{Decomposition of a vertex, generated by the Coulomb operator $K(x-y;A)$, into products
of operators.  In the tree diagram decomposition, $G$ and $K$ operators become ghost and Coulomb propagators
respectively, as explained in the text.}
\label{decomp}
\end{figure}

      In the perturbative expansion of $K$, each of
the operators $M$ contains a single $A$ operator, and in a Coulomb vertex we have to choose one or more of these to connect to external gluons
in the initial or final states.  Let us begin with the case of a Coulomb vertex with a single gluon emerging.  Denote by 
$M^*$ the operator which contains the $A$-field contracting with an external gluon field, and the resulting operator (corresponding to the blob with one gluon line coming out) we'll denote $K^A$.  Then we have
\bea
      K^A &=& {1 \over (-\nabla^2)} \sum_{N=0}^\infty (N+1) \sum_{m=0}^{N-1} M^m M^* M^{N-m-1}
\non \\
&=& {1 \over (-\nabla^2)} \sum_{m=0}^\infty \sum_{n=0}^\infty (m+n+2) M^m M^* M^n
\non \\
&=& {1 \over (-\nabla^2)} \sum_{m=0}^\infty \sum_{n=0}^\infty (m+n+2) M^m (gfA^*\pa){1 \over (-\nabla^2)} M^n
\non \\
&=& {1 \over (-\nabla^2)} \sum_{m=0}^\infty  (m+1) M^m (gfA^*\pa){1 \over (-\nabla^2)}\sum_{n=0}^\infty M^n
\non \\  
& & +  {1 \over (-\nabla^2)} \sum_{m=0}^\infty M^m (gfA^*\pa){1 \over (-\nabla^2)}\sum_{n=0}^\infty (n+1) M^n
\non \\
&=& K (gfA^*\pa) G + G (gfA^*\pa) K  \ .
\eea
Now let's consider a blob with two gluon lines coming out.  By the same reasoning
\begin{widetext}
\bea
K^{AA} &=&   {1 \over (-\nabla^2)} \sum_{N=0}^\infty (N+1) \sum_{m=0}^{N-2} ~~ \sum_{n=0}^{N-m-2}
                         M^m M^* M^n M^* M^{N-m-n-2}
\non \\
            &=&    {1 \over (-\nabla^2)} \sum_{m=0}^\infty \sum_{n=0}^{\infty} \sum_{k=0}^{\infty}
                         (m+n+k+3) M^m M^* M^n M^* M^k                     
\non \\
&=& {1 \over (-\nabla^2)} \sum_{m=0}^\infty \sum_{n=0}^{\infty} \sum_{k=0}^{\infty} (m+n+k+3)
M^m (gfA^*\pa){1 \over (-\nabla^2)} M^n (gfA^*\pa){1 \over (-\nabla^2)} M^k
\non \\
&=& {1 \over (-\nabla^2)} \sum_{m=0}^\infty (m+1) M^m  (gfA^*\pa){1 \over (-\nabla^2)} \sum_{n=0}^\infty
M^n (gfA^*\pa){1 \over (-\nabla^2)} \sum_{k=0}^\infty M^k 
\non \\
& & + {1 \over (-\nabla^2)} \sum_{m=0}^\infty M^m  (gfA^*\pa){1 \over (-\nabla^2)} \sum_{n=0}^\infty (n+1)
M^n (gfA^*\pa){1 \over (-\nabla^2)} \sum_{k=0}^\infty M^k 
\non \\
& &+ {1 \over (-\nabla^2)} \sum_{m=0}^\infty M^m  (gfA^*\pa){1 \over (-\nabla^2)} \sum_{n=0}^\infty
M^n (gfA^*\pa){1 \over (-\nabla^2)} \sum_{k=0}^\infty (k+1) M^k 
\non \\
&=&  K(gfA^*\pa)G(gfA^*\pa)G + G(gfA^*\pa)K(gfA^*\pa)G + G(gfA^*\pa)G(gfA^*\pa)K \ .
\non \\
\label{KAA}
\eea
\end{widetext}
The general case follows from induction, and is indicated schematically in Fig.\ \ref{decomp}.  The tree diagram
approximation is to replace the operators $K,G$ by the corresponding propagators, as shown in the figure.  In diagrammatic terms this means we either neglect completely contractions of field operators in one ($K$ or $G$) blob with field operators in another blob, or we assume that the only effect of such contractions is to multiply the point-like
vertex by a constant, which can be absorbed into a multiplicative factor $c$ in the ghost propagator.

   At this stage it would appear that the Coulomb vertices involve $K$-propagators rather than Coulomb propagators.
However, this neglects the fact that there is a second contribution to vertices with $N$ external gluon lines, which comes about from expanding the $K\rho \rho K$ operator.  We recall that the Coulomb propagator is actually a sum of the K-propagator and $\langle K \rho \rho K \rangle$.  A similar statement is true for Coulomb vertices.  Let us consider one term in the expansion
of the $K$ operator with $N$ external gluon operators, in which there are $n$ products of $G(gfA^*\pa)$ to the left of
the $K$ operator, and $N-n$ products of $(gfA^*\pa)G$ operators to the right of the $K$ operator, i.e.
\bea
    T_1 &=&  \stackrel{\leftarrow n \rightarrow}{G(gfA^*\pa) \ldots G(gfA^*\pa)}K
             \stackrel{\leftarrow N-n \rightarrow}{(gfA^*\pa)G \ldots (gfA^*\pa)G} \ .
\non \\
\eea
Again, in the tree diagrammatic expansion we neglect any diagrams which connect $G$ operators to $K$ operators or to
other $G$ operators, and the result is a product of ghost propagators, $K$-propagators, and $A$ field operators which contract with operators in the initial and/or final states.
To each term of this kind, there is a corresponding term which arises from the expansion of the $K \rho \rho K$ operator:
\bea
T_2 &=& \bigg\{ \stackrel{\leftarrow  n \rightarrow}{G(gfA^*\pa) \ldots G(gfA^*\pa)}K \rho \bigg\}
\non \\
        & &  \times   \bigg\{   \rho K \stackrel{\leftarrow N-n \rightarrow}{(gfA^*\pa)G \ldots (gfA^*\pa)G} \bigg\} \ .
\eea
In the tree diagram expansion the $G$ operators become ghost propagators, and $K\rho \rho K$ becomes 
$\langle K\rho \rho K \rangle$.  Then in the tree-level approximation to $T_1+T_2$ the $K$ operators appear in the
combination $\langle K \rangle + \langle K\rho \rho K \rangle$, which is simply the Coulomb propagator $\tK$.  This is completely
general.  The effect of adding the $K\rho \rho K$ operator in computing Coulomb vertices in this framework is equivalent to
considering only the expansion of the $K$ operator, and then replacing the $K$-propagator with the Coulomb propagator.

    One might ask why we have neglected terms in the expansion of $ K\rho \rho K  $ such as
\bea
 \bigg\{G(gfA^*\pa) \ldots G(gfA^*\pa) K(gfA^*\pa)G \ldots (gfA^*\pa)G \rho \bigg\}
\non \\
          \times    \bigg\{\rho G(gfA^*\pa) \ldots G(gfA^*\pa) K(gfA^*\pa)G \ldots (gfA^*\pa)G  \bigg\}  \ .
\non \\
\eea
The answer is that  all of the diagrams which contribute to the VEV of such a quantity involve loops which cannot be
absorbed into a ghost or Coulomb propagator.  In the tree diagram framework we propose here, contributions of that
kind are dropped.

\section{\label{matrix} The Hamiltonian in a basis of zero and one constituent gluons}

    In the toy model of a gluon chain, explained in ref.\ \cite{Greensite:2014bua}, the average number of gluons in the chain will grow linearly with the quark-antiquark separation.   If this idea really works, then there should then be some intermediate distance range, presumably just after the onset of confinement, where the average number of gluons is less than one, i.e.\ the minimal energy state is mostly a superposition of just zero and one-gluon states.  We will denote a finite
set of such states, containing also a static quark-antiquark pair of separation $R$, as $\{ |n \rangle \}$, where $|0\rangle$ is the zero-gluon state
\beq
    |0 \rangle  =  \overline{q}^{\dg \a}(0) q^{\dg \a}(\vR) |\mbox{0}\rangle_{vac} \ ,
\eeq
and the remaining one-gluon states ($n \ge 1$) have the form
\bea
    |n \rangle  &=&   \overline{q}^{\dg \a}(0) \left[ \int d^3x \Psi^{(n)}_i(x) A_i^a(x) t_a^{\a \b} \right] q^{\dg \a}(\vR)         |\mbox{0}\rangle_{vac} \ ,
\non \\
\eea
with $t_a=\oh \l_a$ the SU(3) group generators.  From these states we can construct an orthonormal basis, and
diagonalize the Hamiltonian in that basis.  For this purpose we need to compute 
\beq
       H_{mn} = \langle m | H| n \rangle  ~~~,~~~ O_{mn} = \langle m | n \rangle
\label{HOmn}
\eeq
The trial one-gluon wavefunctions $\Psi^{(n)}_i(x)$ may depend on some variational parameters, which are chosen
such that the lowest energy state obtained by diagonalization is minimized.

    There is a big simplification if we choose our trial one-gluon wavefunctions $\Psi^{(n)}_i(x)$ to be transverse,
$\nabla \cdot \vec{\Psi}^{(n)}=0$.  Then, since the transverse gluon propagator will always contract with the index of
an external wavefunction, and the coordinate $x$ of the external wavefunction is always integrated over, we can simply drop the $k_i k_j/k^2$ piece of the transverse projection operator, because that piece will always act like a divergence on 
$\vec{\Psi}$.  Effectively, then, we are allowed to drop $D^{(1)}_{ij}(x,t)$ in \rf{gluonprop}, and just use
\bea
D_{ij}(x,t) &\ra& \d_{ij} D(x,t) \ ,
\non \\
D(x,t) &=&  \int {d^3k \over (2\pi)^3} { 1 \over 2\o_k} e^{i \vk \cdot \vx} e^{-\o_k t} \ .
\eea
Likewise, for $\e \approx 0$,
\beq
\tD_{ij}(x,\e) \ra {i\over 2} \d_{ij}\mbox{sign}(\e) \d^3(x) \ .
\label{EA}
\eeq
We also denote, at equal-times,  ${D(x) \equiv D(x,t=0)}$.  With transverse $\Psi^{(n)}_i(x)$ and $n\ge 1$
\bea
    O_{mn} &=& \langle m|n \rangle
\non \\
       &=&   \tr t_a t_a \int d^3z_1 d^3z_2 ~ \Psi^{(m)*}_i(z_1) D(z_1-z_2) \Psi^{(n)}_i(z_2) \ .
\non \\
\label{Omn}
\eea
with $\tr (t_a t_a)=3C_F$ where $C_F$ is the quadratic Casimir in the fundamental representation.

  Now, for any choice of transverse $\Psi^{(n)}_i(x)$, there exists another transverse function $f^{(n)}_i(x)$, with Fourier transform $f^{(n)}_i(k)$ defined by
\beq
            \Psi^{(n)}_i(x) = \int {d^3k \over (2\pi)^3} \sqrt{2\o_k} f^{(n)}_i(k) e^{ik\cdot x} \ .
\label{Pf}
\eeq
Then
\bea
 \langle n|n \rangle &=& 3 C_F  \int d^3x ~ f^{*(n)}_i(x) f^{(n)}_i(x)  \ .
\eea
So $f^{(n)}_i(x)$ is the analog of an ordinary quantum mechanics one-particle wavefunction, and ideally we would like to first choose $f^{(n)}_i(x)$ based on some combination of intuition and convenience, transform to $f^{(n)}_i(k)$, and then compute $\Psi^{(n)}_i$ from \rf{Pf}.  But in practice it would be difficult to derive $\Psi^{(n)}_i$ analytically, because of the 
$\sqrt{\o_k}$ factor in \rf{Pf}.  Even numerically this would be challenging, because the integrand of \rf{Pf} is an oscillating function.  It is simpler to work directly with $\Psi^{(n)}_i$. 

\subsection{The truncated basis}

   For the numerical work in this article, we will choose as an ansatz transverse wavefunctions of the form
\bea
 \vec{\Psi}^{(n)} &=& \nabla \times \left[ \begin{array}{c}
                               -y \cr
                               x \cr
                               0 \end{array} \right] F_{\a \b}(x,y,z)
\non \\
&=&  \left[ \begin{array}{c}
                               -x \pa_z F_{\a \b} \cr
                               -y \pa_z F_{\a \b} \cr
                               2F_{\a \b} + x\pa_x F_{\a \b} + y \pa_y F_{\a \b} \end{array} \right] \ ,
\eea
with $\a=1,2,..,N, ~ \b=1,2,..,M$ and $n=(\a-1)M + \b$.  Then $F_{\a \b}$ is chosen to have the form
\begin{widetext}
\beq
      F_{\a \b}(x,y,z) = f_\a(z) L_{\b-1}^1\left({4r \over a}\right) 
           \exp\left[-{1\over a}\left(\sqrt{r^2 + z^2} + \sqrt{r^2 + (R-z)^2} \right)\right]   \ ,
\label{Fnm}
\eeq
\end{widetext}
where $r=\sqrt{x^2 +y^2}$ is the polar coordinate in the transverse direction, $L_\b^1$ is an associated Laguerre polynomial, $a$ is a variational parameter, and
\beq
     f_\a(z) = \left\{ \begin{array}{cl}
                       1 & \a=1 \cr
                          &        \cr
                 \sin\left({\pi \a \over R+2z_0}(z+z_0)\right) & \a >1  \end{array} \right.   \ .
\eeq
The motivation for this choice of $F_{\a\b}$ is the fact that if a gluon is located on the $z$-axis between the quarks, with
$x=y=0$ and $0 < z < R$, then the linear piece of the Coulomb potential between the gluon, the quark, and the antiquark, sums up to $\s_c z + \s_c (R-z) = \s_c R$ which is
independent of $z$.  So if the gluon is not too far away from the $z$-axis, it can move
more or less freely in the $z$-direction between the quarks.  Hence the excitations in the $z$-direction might be
as in a one-dimensional square well of width $ \approx R$.  Since there is no reason for the wavefunction to drop to zero exactly
at $z=0,R$, we allow for a ``well'' a little bigger than $R$, namely $R+2z_0$, where in this work we have taken $z_0=a/3$. 
Note that for $r \gg |z|, |R-z|$
\beq
           F(x,y,z) \sim \exp(-2r/a) \ .
\eeq
The set of orthogonal functions in the integration measure $ r e^{-4r/a} $
are the associated Laguerre polynomials $L_m^1(4r/a)$. 

   It should be understood that an orthogonal set of $F_{\a \b}$ does not produce an orthogonal
set of one gluon states $\{ |n \rangle \}$; it only specifies a set of linearly independent states which together span a
subspace of Hilbert space.  The object is to find the lowest eigenvalue of the Hamiltonian in this subspace.
If we define a matrix $[H]$ with elements $H_{mn}$ and a corresponding matrix $[O]$ of overlaps $O_{mn}$ in the non-orthogonal set of states $\{ |n \rangle \}$, as in \rf{HOmn}, the eigenvalues of the Hamiltonian operator restricted to this subspace are obtained by solving the generalized eigenvalue equation
\beq
             [H] \vec{u}^{(n)} = E_n [O] \vec{u}^{(n)}
\eeq 
This is essentially equivalent to the alternate procedure of constructing a set of orthonormal states 
$\{|v_n \rangle \}$ from the $\{ |n \rangle \}$
via, e.g., the Gram-Schmidt procedure, computing matrix elements $H'_{ij}=\langle v_i|H| v_j \rangle$ in the orthogonal basis from the $H_{mn}$, and then diagonalizing the matrix $H'$ in the usual way.  

\subsection{Matrix Elements}

We now list the improved-tree contributions to each of the required Hamiltonian matrix elements $H_{mn}$ of eq.\ \rf{HOmn}.  The state with a static quark-antiquark pair and zero constituent gluons will be denoted $|0\rangle$, and
states with the static quarks and one consituent gluon will be denoted $|n \rangle$ with $n>0$.  The expression for overlap matrix elements $O_{mn}$ of one-gluon states was given in \rf{Omn}, while
for the zero-gluon state $O_{00} = 3$ (a sum over three colors), and $O_{0n}=0$. \\

\bigskip

\ni \underline{1. ~ $\langle 0|H|0 \rangle$} \\

\begin{figure}[h!]
\centerline{\scalebox{0.35}{\includegraphics{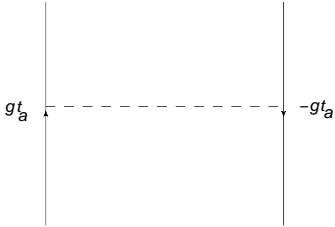}}}
\caption{Diagram responsible for the instantaneous Coulomb potential.}
\label{D00}
\end{figure}

The matrix element is given by the diagram in Fig.\ \ref{D00}, and this is the simplest contribution. It is
\bea
      H_{00}  = -g^2 \tr (t_a t_a) \tK(R) = 3 K_c(R) \ .
\label{H00}
\eea

\ni \underline{2. ~ $\langle m|H|n \rangle$ kinetic term} \\

\begin{figure}[h!]
\centerline{\scalebox{0.35}{\includegraphics{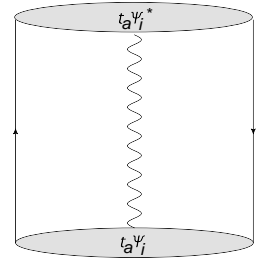}}}
\caption{Diagram associated with constituent gluon kinetic energy.}
\label{D11kin}
\end{figure}

    The next simplest term is associated with the kinetic energy of the one-gluon state for the case $m=n$, and that is best obtained by displacing the initial and final state in time, so that the transverse gluon propagator is at unequal times, and then taking minus the time derivative of the diagram shown in Fig.\ \ref{D11kin}:
\bea
    H^{kin}_{mn} &=& -\lim_{t \ra 0} {d\over dt}   \tr t_a t_a \int d^3z_1 d^3z_2 ~ \Psi^{(m)}_i(\vz_1) D(\vz_1-\vz_2,t) 
                                \Psi^{(n)}_i(\vz_2)
\non \\
&=&   -  \tr t_a t_a  \int d^3z_1 d^3z_2 ~  \Psi^{(m)}_i(\vz_1) D'(\vz_1-\vz_2,0) \Psi^{(n)}_i(\vz_2)
\non \\
&=&   3 C_F  \int d^3z  ~ \oh \Psi^{(n)}_i(\vz) \Psi^{(n)}_i(\vz)  \ .
\label{H11kin}
\eea
From \rf{Pf}, this could also be expressed in terms of $f_i$ as
\beq
H^{kin}_{mn} \propto \int d^3k ~ \o_k f^*_i(k) f_i(k) \ ,
\eeq
although this is actually not the most useful form if the variational wavefunction is given in terms of $\Psi_i$ 
rather than $f_i$.

\bigskip
\ni \underline{3. ~ $\langle n|H|0 \rangle$} \\

\begin{figure}[t!]
\centerline{\scalebox{0.55}{\includegraphics{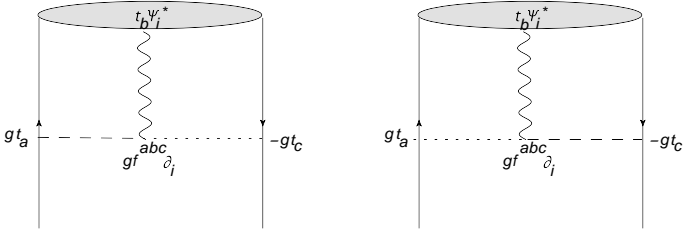}}}
\caption{Diagram responsible for the $H_{n0}$ Hamiltonian matrix element.}
\label{D01}
\end{figure}

   The relevant diagrams are shown in Fig.\ \ref{D01}, and we find
\bea
            H_{n0} &=&  -2i {C_A \over C_F} \int d^3x d^3z ~ \Psi^{(n)}_i(\vz) D(\vx-\vz) 
\non \\
& & \qquad \times  \big\{ G_c(\vx) \pa_i K_c(\vR-\vx) + K_c(\vx) \pa_i G_c(\vR-\vx) \big\}
\non \\
\label{H10}
\eea

\bigskip

\ni \underline{4. ~ $\langle m|H|n \rangle$ Coulomb term} \\

\begin{figure}[htb]
\centerline{\scalebox{0.55}{\includegraphics{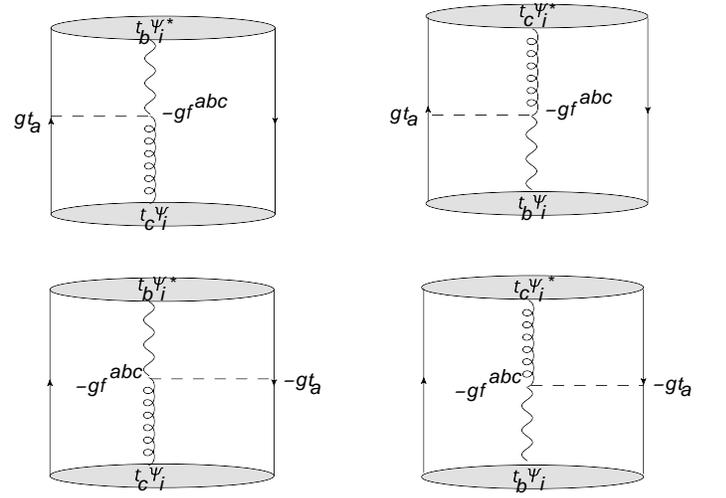}}}
\caption{Planar diagrams contributing to the instantaneous Coulomb energy of the one constituent gluon state.}
\label{D11coul}
\end{figure}

   The planar diagrams are shown in Fig.\ \ref{D11coul}.  There are two diagrams in which the Coulomb propagator ends on
the left hand side quark, and likewise two diagrams where the Coulomb propagator ends on
the right hand side antiquark.  The sum of all contributions is 
\begin{widetext}
\bea   
H_{mn}^{coul} &=&  {C_A \over C_F} \int d^3z_1 d^3z_2 ~ \Psi_i^{(m)}(\vz_1) D(\vz_1-\vz_2) \Psi_i^{(n)}(\vz_2)
 \big\{ K_c(\vz_1) + K_c(\vR - \vz_1) +  K_c(\vz_2) + K_c(\vR - \vz_2) \big\} \ .
\label{H11coul}
\eea
\end{widetext}

\begin{figure}[htb]
\centerline{\scalebox{0.3}{\includegraphics{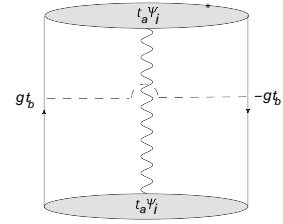}}}
\caption{Non-planar contribution to the instantaneous Coulomb energy of the one constituent gluon state.}
\label{D11np}
\end{figure}

   There is also a non-planar contribution to the Coulomb energy, which is shown in Fig.\ \ref{D11np}.  This contribution
comes out to be
\beq
         H^{np}_{mn} = -{1\over 8} K_c(R) O_{mn} \ .
\label{H11np}
\eeq

\ni \underline{5. ~ $\langle m|H|n \rangle$ $KGG$ terms} \\

\begin{figure}[t!]
\subfigure[]  
{   
 \label{D11a}
 \includegraphics[scale=0.35]{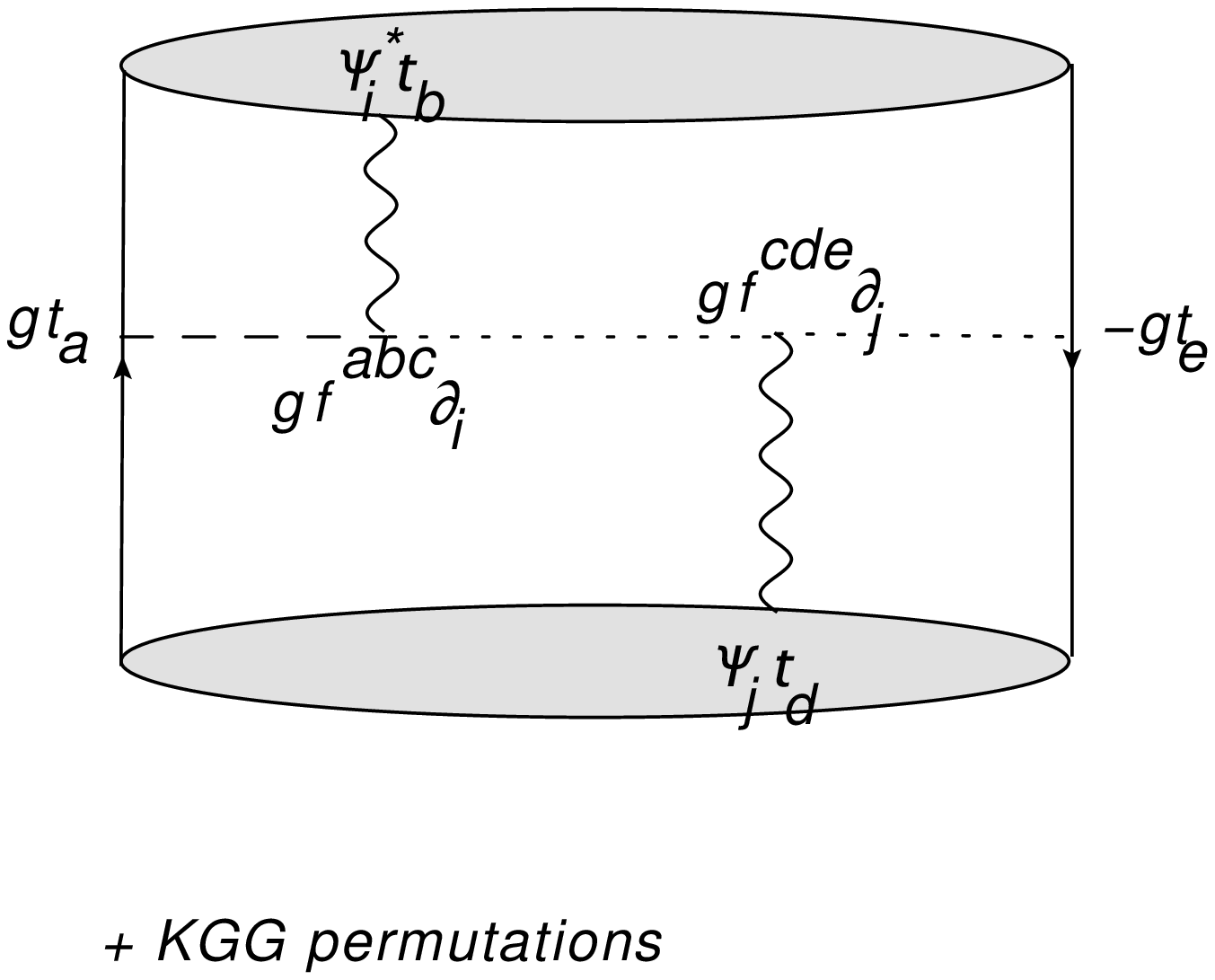}
}
\subfigure[]  
{   
 \label{D11b}
 \includegraphics[scale=0.35]{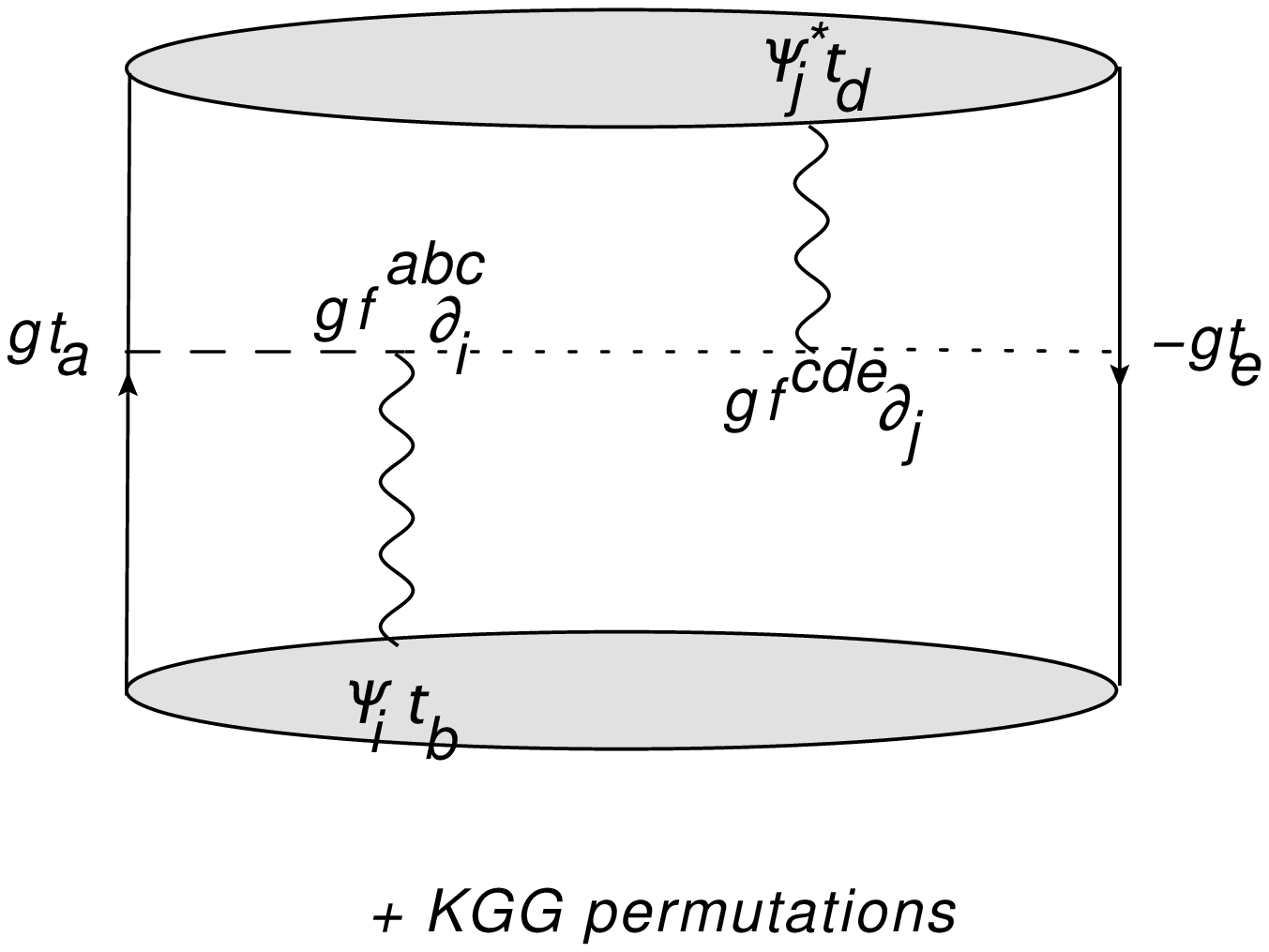}
}
\caption{Diagrams contributing to the $H^{KGG}$ matrix element.  These entail a 12-dimensional integration.} 
\end{figure}

   The diagrams are shown in Figs.\ \ref{D11a} and \ref{D11b}.  ``KGG permutations'' means permutations in the
order of the instantaneous Coulomb (K) and ghost (G) propagators.  The possible orders are KGG, GKG, and GGK.
With the diagrammatic rules as before (trace of matrix generators is clockwise), it turns out that diagram \ref{D11b} is
simply the complex conjugate of diagram \ref{D11a}.  The resulting contribution is

\begin{widetext}
\bea
\lefteqn{H_{mn}^{KGG} =}
\non \\
&=&  {C_A^2 \over C_F} \int d^3x d^3y d^3u d^3v ~ D(\vu) D(\vv) \bigg\{ \Psi_i^{(m)}(\vx-\vu) \Psi_j^{(n)}(\vy-\vv) 
+  \Psi_i^{(n)}(\vx-\vu) \Psi_j^{(m)}(\vy-\vv)   \bigg\}     
\non \\   
& & \times \bigg\{ K_c(\vx) \pa^x_i G_c(\vx-\vy) \pa^y_j G_c(\vy-\vR) + G_c(\vx) \pa^x_i K_c(\vx-\vy) \pa^y_j G_c(\vy-\vR) + 
G_c(\vx) \pa^x_i G_c(\vx-\vy) \pa^y_j K_c(\vy-\vR) \bigg\}
\label{H11KGG}
\eea
\end{widetext}
 
\section{\label{gluons}The equal-times transverse gluon propagator}

   In order to actually compute the $H_{mn}$ matrix elements we need the full propagators
$K_c(x-y)$, $G_c(x-y)$, and $D(x-y)$.  The Coulomb propagator follows from \rf{VK} and the Coulomb potential determined previously in 
\cite{Greensite:2014bua}.  In this section we present our lattice Monte Carlo results for $D(x-y)$; the infrared behavior
of the ghost propagator will be taken from refs.\ \cite{Burgio:2008jr,Burgio:2012bk,Nakagawa:2009zf}.

\subsection{The question of gauge copies}

    Since our calculation is in Coulomb gauge, we will begin with some remarks about Gribov copies.  It is well known that any point in field space where the quantity
\beq
           R = {1\over 3V_3} \sum_{x} \sum_{i=1}^3 \text{ReTr}[U_i(x,t)] \ ,
\eeq
evaluated on each timeslice, is stationary ($V_3$ is the number of sites on the timeslice), will satisfy the Coulomb gauge condition
\beq
             \sum_{k=1}^3 (U_k(x,t) - U_k(x-\hat{k},t) - \text{h.c.} = 0
\eeq
at all sites $x$, which is just the lattice version of $\nabla \cdot A=0$.  Gribov copies are the configurations on a gauge orbit  which satisfy this condition, and those copies which are local maxima of $R$ are said to be inside the Gribov region.  The global maximum of $R$ is said to be in the ``fundamental modular region.''  In practice it is impossible to find the global maximum of the gauge fixing condition, although some studies use simulated annealing in an attempt to find a better gauge
copy than what might be obtained from, e.g., the ordinary over-relaxation technique.  Here we should remark that in 
a fundamental sense there is no such thing as a ``better'' gauge copy.  It is obvious that if we are calculating a gauge-invariant quantity, then the choice of gauge copy is irrelevant.  Of course, it may be that gauge dependent quantities such 
as propagators and vertices will vary somewhat from copy to copy, and the expectation value for such quantities
could very well depend on exactly how the gauge copy is chosen.  On the other hand, if one could put together propagators and vertices so as to compute a gauge-invariant quantity (such as a scattering amplitude in ordinary perturbation theory, or the energy of some physical state), then it is expected that the gauge variance of the propagators and vertices should cancel out, and the details of Coulomb gauge fixing should be irrelevant.  For this to happen, however, it is important that the same version of Coulomb gauge fixing is used for all gauge-dependent quantities.

    Any lattice Monte Carlo calculation in Coulomb or Landau gauge uses some version of what might be called ``computer gauge.''  Lattice configurations are generated according to the gauge-invariant probability weighting, and then some
deterministic procedure is applied to arrive at a particular Gribov copy inside the Gribov region.  The procedure we
have used in \cite{Greensite:2014bua}, and that we will continue to use here, is the method of Fourier acceleration, introduced in \cite{Davies:1987vs}.  Although we will not use simulated annealing to try to obtain ``better'' gauge copies, it may still be of interest to compare values of $R$ in Coulomb gauge obtained from simulated annealing followed by
Fourier acceleration, with the values obtained from Fourier acceleration alone.  We have carried out these procedures
on the $t=1$ timeslice of 25 lattices of volume $24^4$, generated at gauge coupling $\b=6.0$, with each lattice separated by 2000 update sweeps.  Each timeslice was first fixed to Coulomb gauge by the Fourier acceleration method (which stops after a certain convergence criterion is met), and the value of $R$ (which is approximately 0.88 at this value of $\b$) was recorded.  The timeslice was then subject to a random gauge transformation followed by 1000 simulated annealing sweeps, with the final fixing carried out via Fourier acceleration, and the procedure again stops when the convergence criterion is met. In every case, the values of $R$ obtained for each lattice by the two procedures differed at most at the fourth non-zero digit, i.e.\ a difference on the order of $10^{-4}$.  A variety of cooling schedules and final temperatures  were tried out for the simulated annealing steps, and seemed to make little difference to this result.  It may be that Fourier acceleration is already very efficient at maximizing $R$.

\subsection{Renormalization}

   Propagators computed in lattice simulations are bare propagators.  The relation of the bare to the renormalized   transverse gluon propagator is given by the usual relation
\beq
          D^{bare}_{ij}(x) = Z_A^2 D_{ij}(x) \ ,
\eeq
where $Z_A$ is the wavefunction renormalization factor, which depends on both the cutoff and the renormalization scheme. Relations between various renormalization constants in Coulomb gauge were
worked out in \cite{Zwanziger:1998ez}, where we find that $g^2 \tK(x)$ in terms of bare coupling and propagator equals the same expression in terms of the renormalized coupling and propagator.  From this reference one can also deduce that
the combination $g G(x) A_k$ is renormalization invariant, in which case
\beq
          g_{bare} G_{bare}(x)  =   g G(x)  Z_A^{-1} \ .
\eeq
It is worth noting that Coulomb vertices, in Hamiltonian matrix elements, are always multiplied by a factor of $g^2$.  This means that the tree level Coulomb vertices discussed in subsection involve only renormalization group invariant combinations $g^2 \tK(x-y)$ and $g G(x-y) A(y)$.

   This observation has an important consequence:  it means that when states are normalized (i.e.\ by division by 
$\sqrt{O_{nn}}$), the Hamiltonian matrix elements we compute are renormalization group invariant.  Therefore if we compute ghost, gluon, and Coulomb propagators at a particular lattice coupling $\b$, and use these to calculate the energy spectrum of the static quark-antiquark pair plus constituent gluons, the continuum limit obtained by extrapolation to 
$\b \ra \infty$ will be the same as if we had carried through the spectrum calculation using instead the corresponding renormalized propagators and vertices.

\subsection{Numerical results}

   We define the space components of $A$-field on the lattice, in Coulomb gauge, as 
\beq
          A_i(\vx,t) = {1\over 2 i g a} \big( U_i(\vx,t) - U_i^\dg(\vx,t)) \ ,
\eeq
where $g=\sqrt{6/\b}$ and $a=a(\b)$ is the lattice spacing in fermis at coupling $\b$, obtained from the Necco-Sommer
formula \cite{Necco:2001xg}
\bea
a(\b) &=& (0.5 \mbox{~fm}) ~ \exp\big[-1.6804 - 1.7331 (\b - 6) 
\non \\
& & \qquad + 0.7849 (\b - 6)^2 - 0.4428 (\b - 6)^3] \ .
\label{Necco}
\eea
We then use lattice Monte Carlo simulation (with Coulomb gauge fixing as described above) to compute the equal-times expectation value
\beq
\D(R) = {1\over 3} \d_{ij} \langle \tr[A_i(\vx,t) A_j(\vy,t)] \rangle \ ,
\eeq
where $R=|\vx-\vy|$.\footnote{Note that a mid-link prescription for position would make no difference to separation $R$, given the $i=j$ restriction.}.  In terms of the propagator 
\bea
D_{ij}^{ab}(R) &=& \langle A^a_i(\vx,t) A_j^b(\vy,t)] \rangle
\non \\
                       &=& \d^{ab} D_{ij}(R)
\eea
we have
\bea
\D(R) &=& {1\over 3} \d_{ij} \oh \d^{ab} D_{ij}^{ab}(R) = {4\over 3} \d_{ij} D_{ij}(R)
\non \\
&=& {8\over 3} D(R) \ .
\label{D}
\eea

\begin{figure*}[t!]
\subfigure[~$\b=5.7$]  
{   
 \label{logb57}
 \includegraphics[scale=0.6]{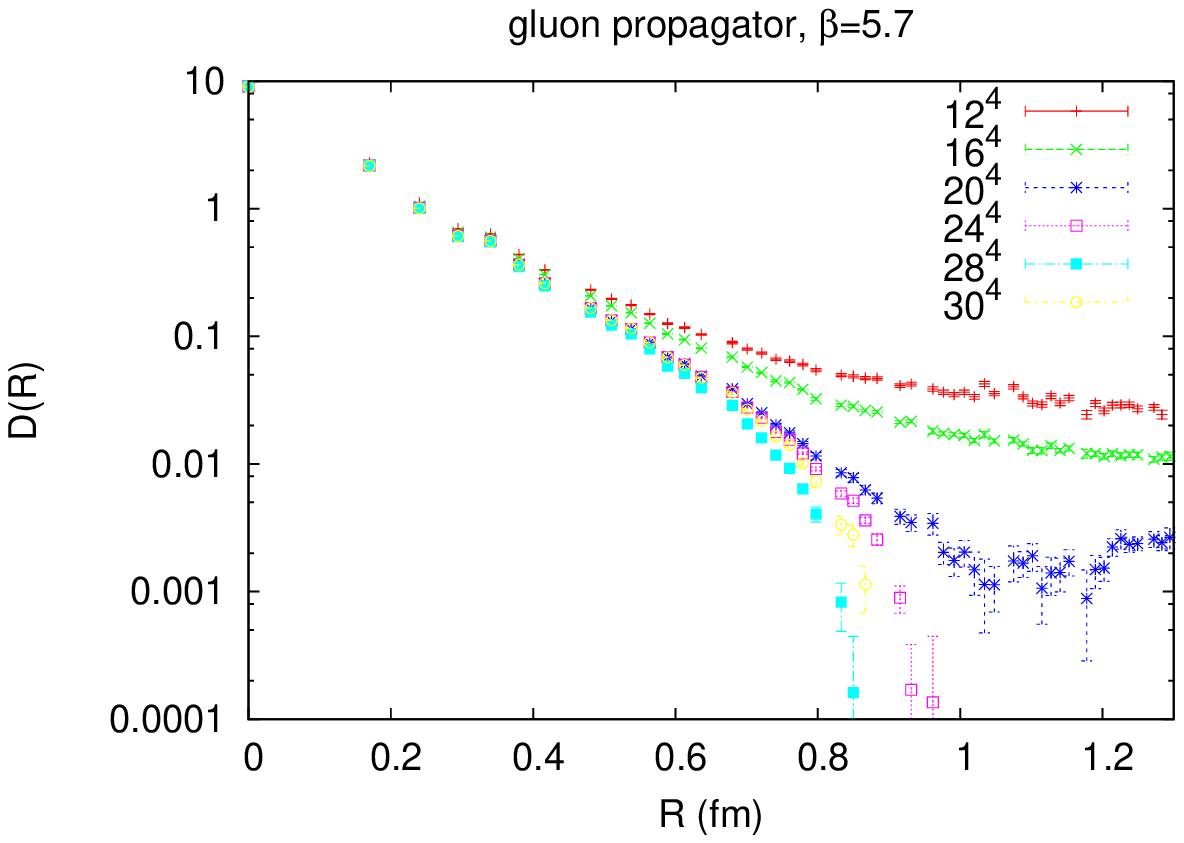}
}
\subfigure[~$\b=5.8$]  
{   
 \label{logb58}
 \includegraphics[scale=0.6]{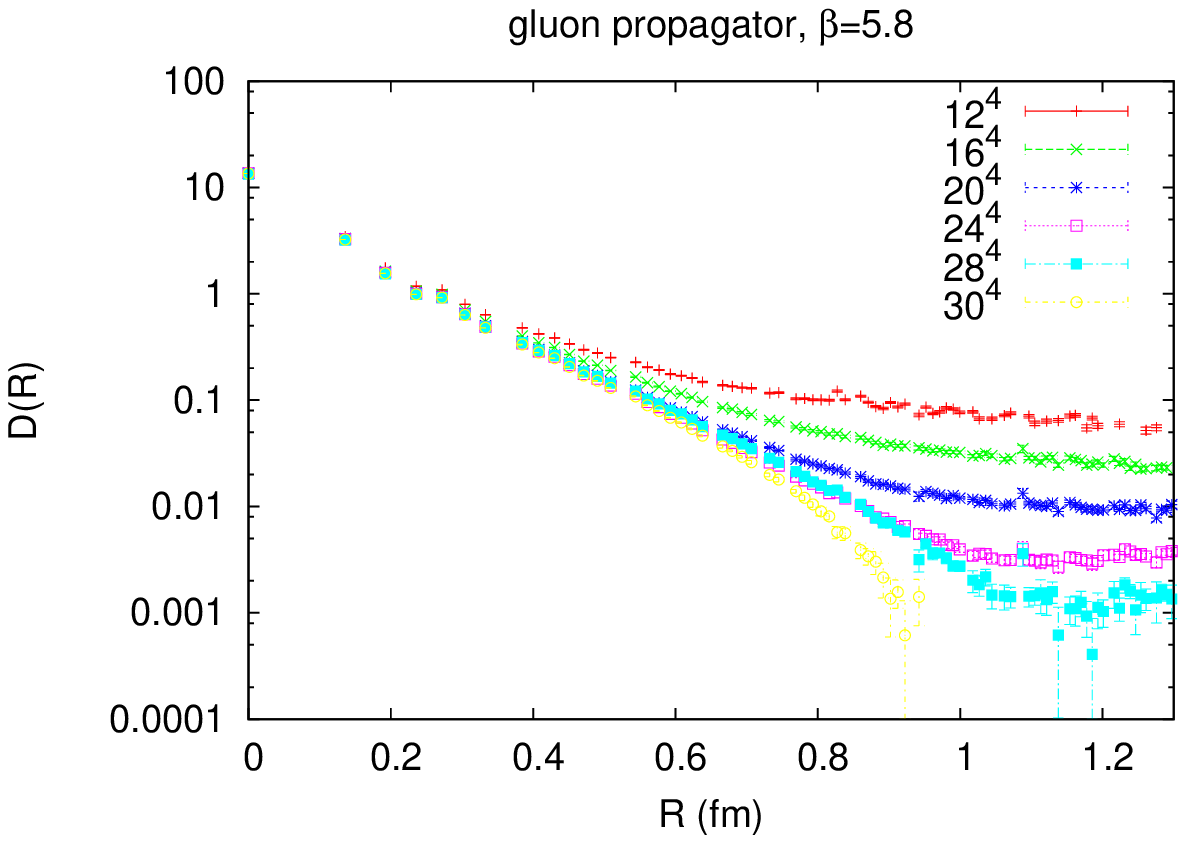}
}
\subfigure[~$\b=5.9$]  
{   
 \label{logb59}
 \includegraphics[scale=0.6]{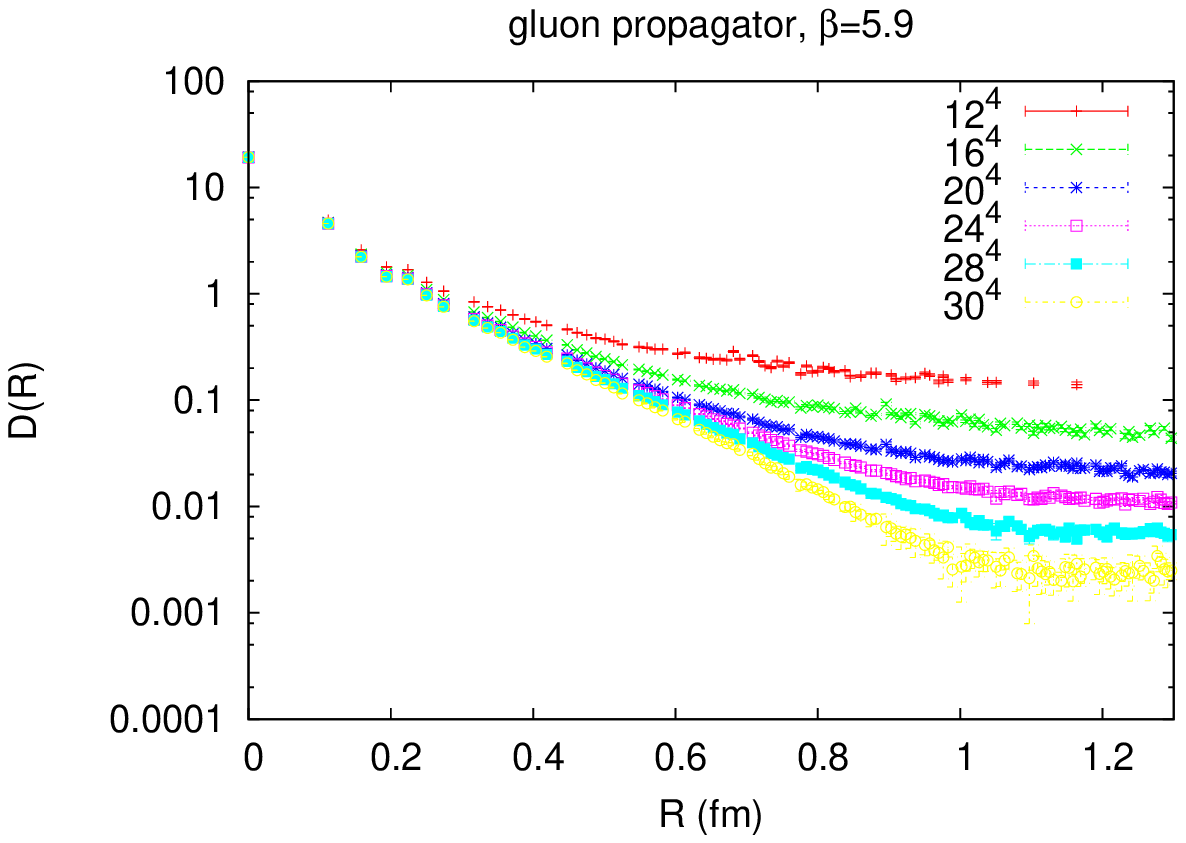}
}
\subfigure[~$\b=6.0$]  
{   
 \label{logb60}
 \includegraphics[scale=0.6]{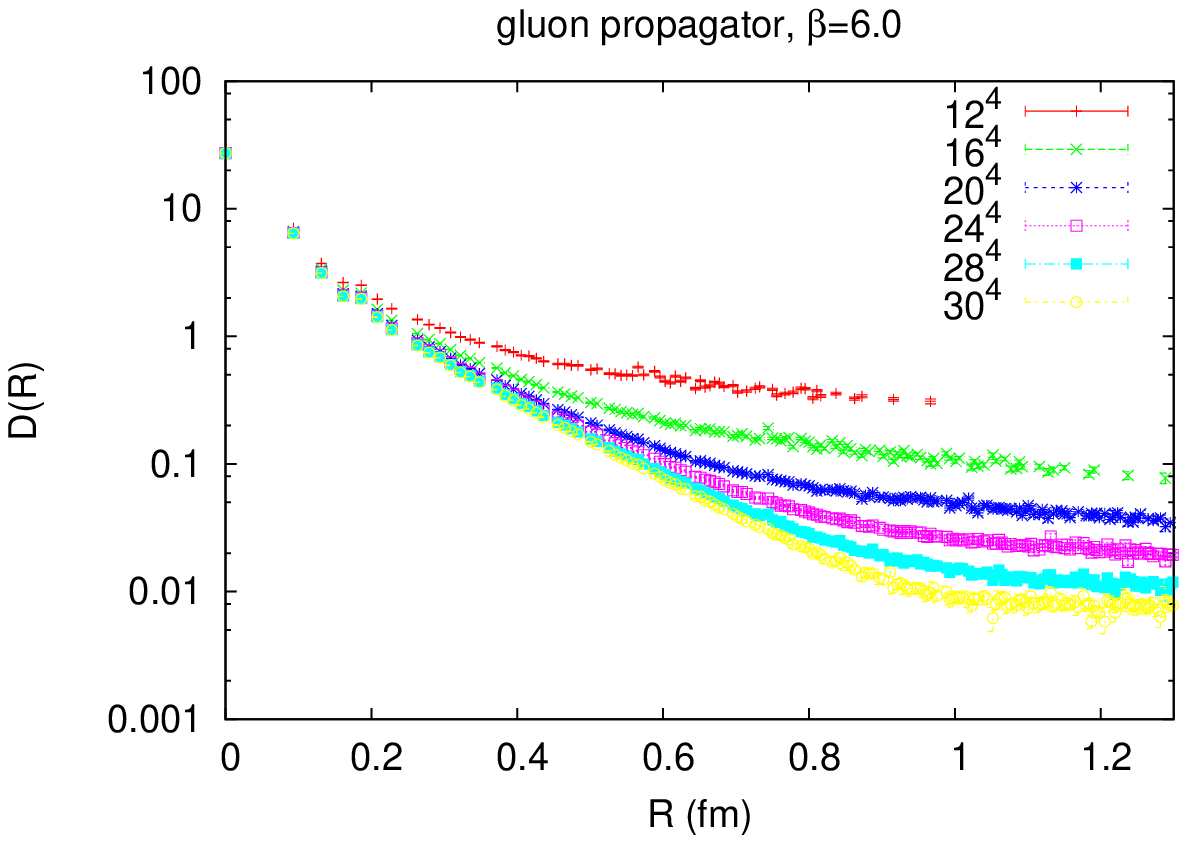}
}
\caption{Lattice Monte Carlo results, in physical units, for the equal-times transverse gluon propagator at lattice volumes
$12^4 - 30^4$ and couplings $5.7\le \b \le 6.0$.  Data is shown on a logarithmic scale.  }
\label{logybeta}
\end{figure*}

\begin{figure*}[t!]
\subfigure[~$\b=5.7$]  
{   
 \label{linb57}
 \includegraphics[scale=0.6]{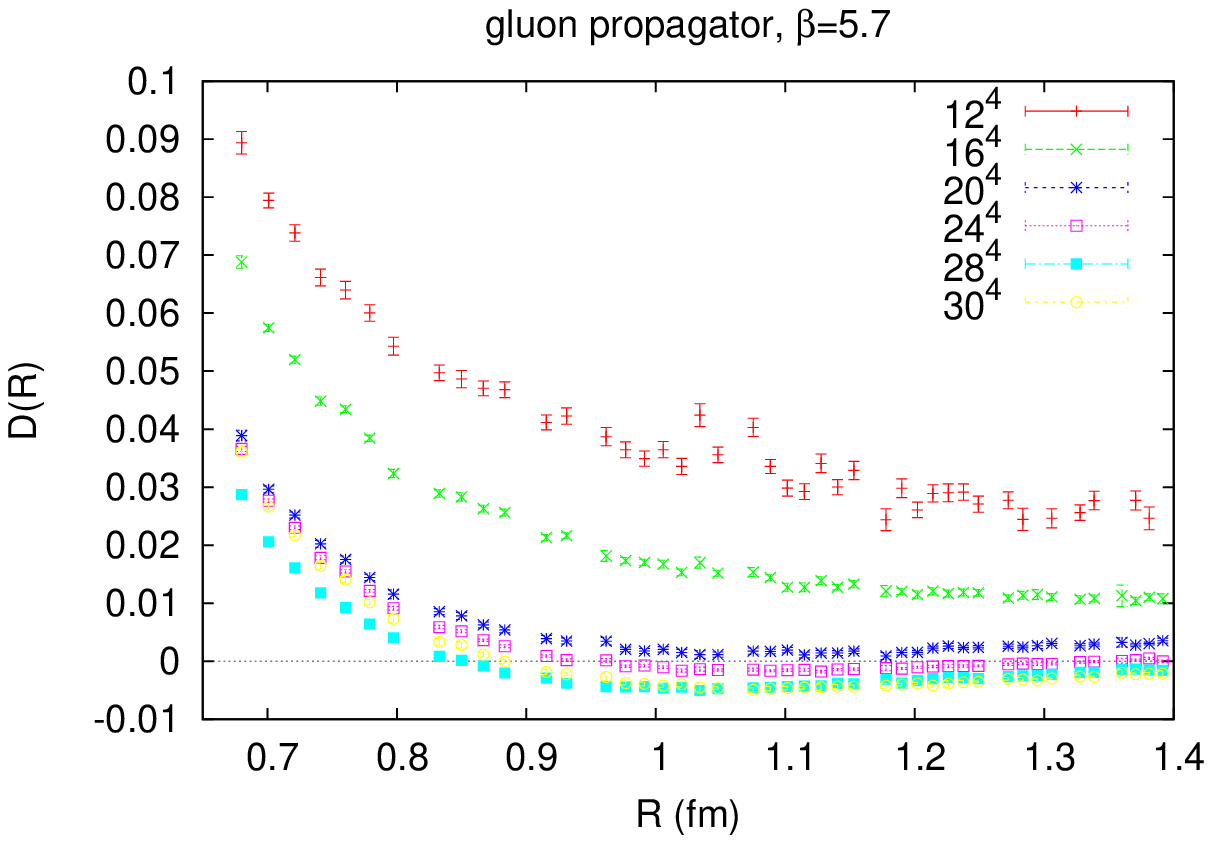}
}
\subfigure[~$\b=5.8$]  
{   
 \label{linb58}
 \includegraphics[scale=0.6]{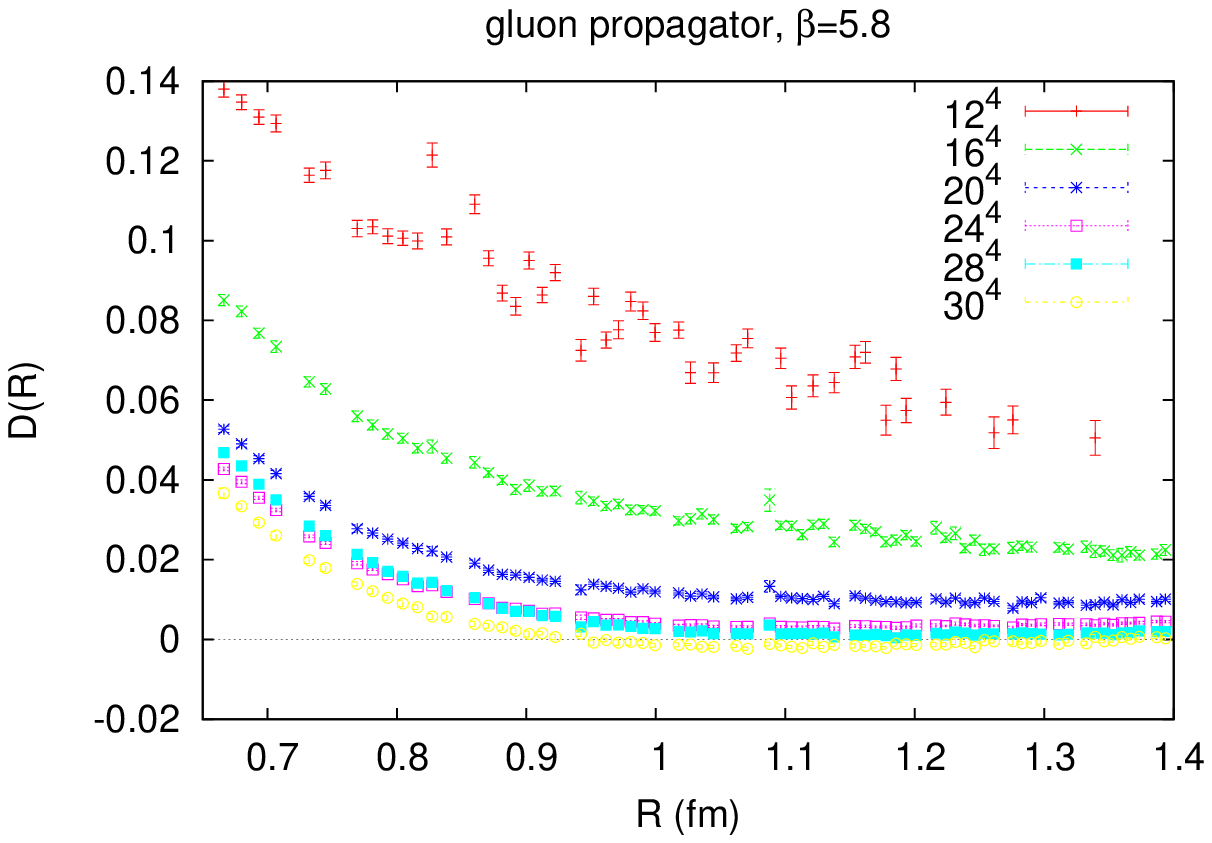}
}
\subfigure[~$\b=5.9$]  
{   
 \label{linb59}
 \includegraphics[scale=0.6]{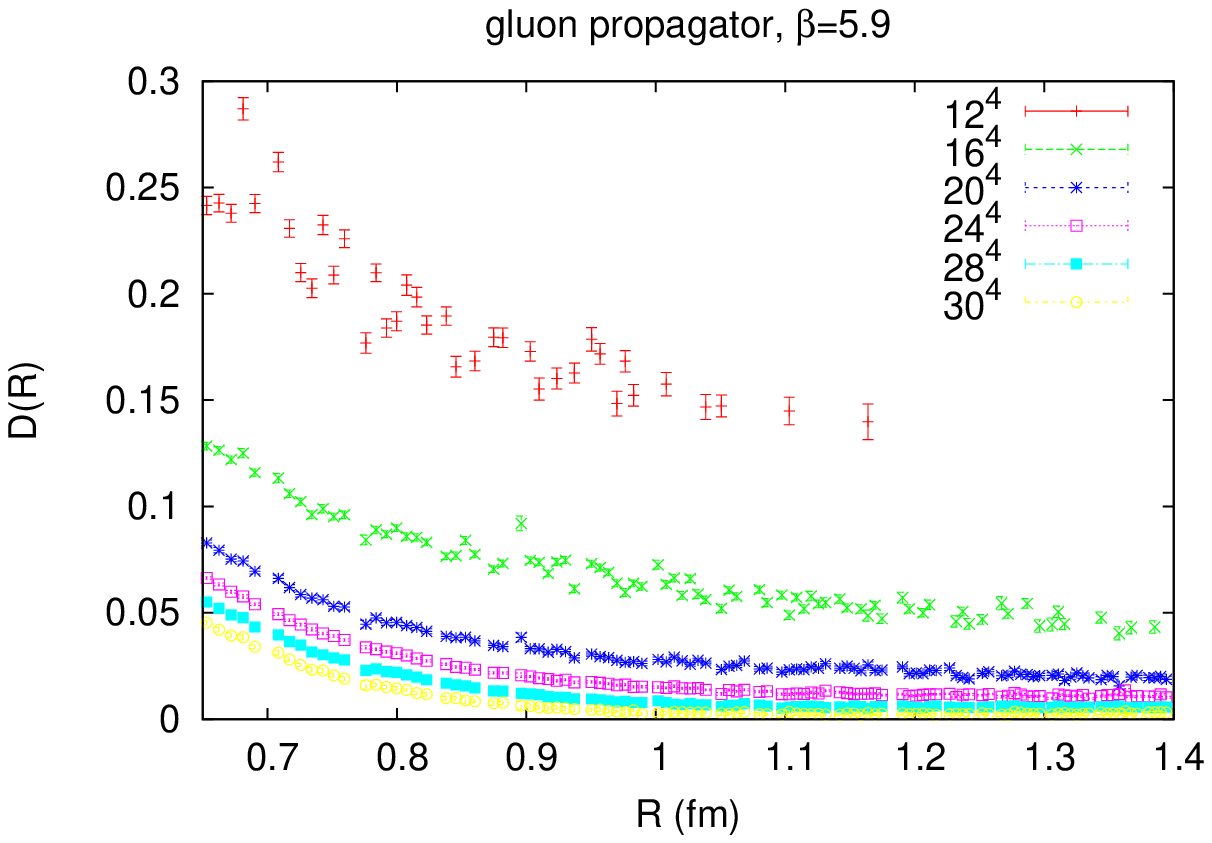}
}
\subfigure[~$\b=6.0$]  
{   
 \label{linb60}
 \includegraphics[scale=0.6]{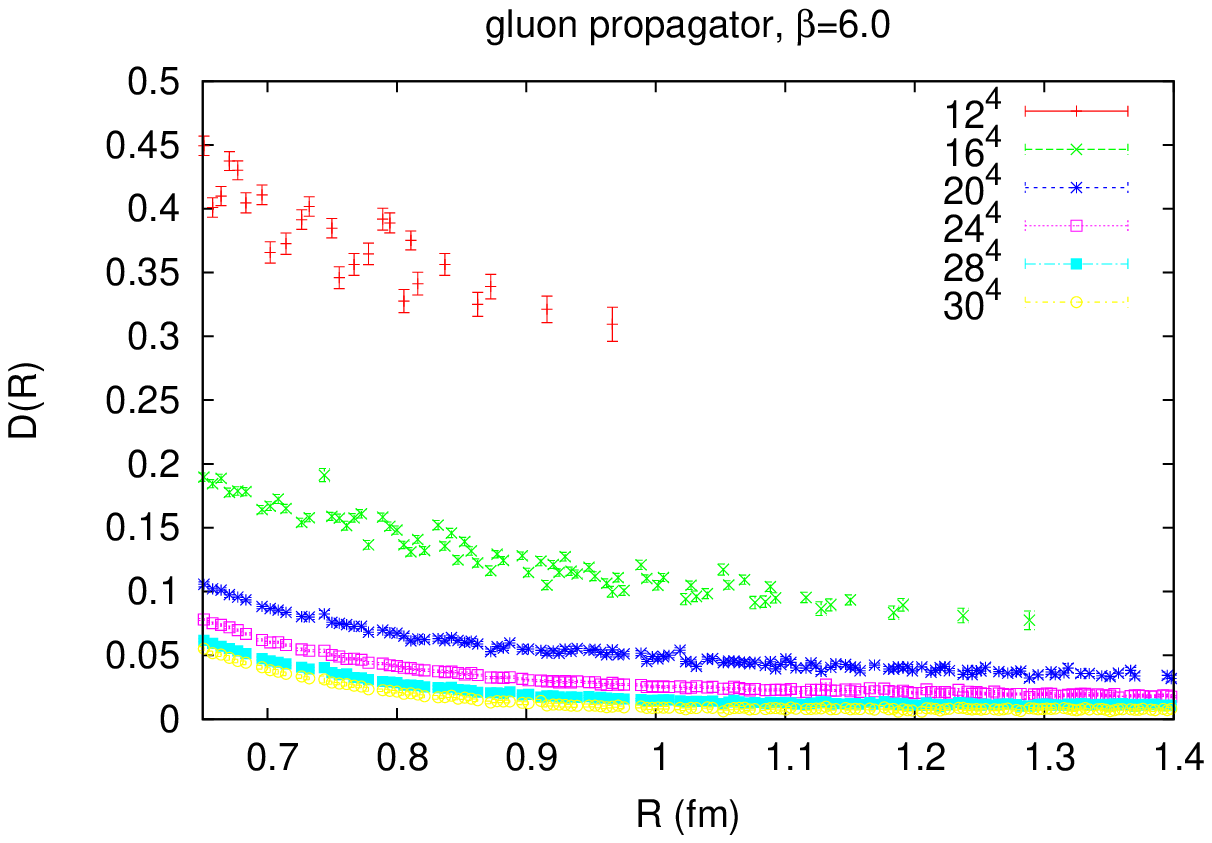}
}
\caption{Same as Fig.\ \ref{logybeta}, but on a linear scale.}
\label{linybeta}
\end{figure*}

    The simulations have been carried out for lattice volumes $12^4, 16^4, 20^4, 24^4, 28^4, 30^4$, and 
 $\b=5.7, 5.8, 5.9, 6.0$; there is roughly a factor of two between lattice spacings at the lowest and highest values of $\b$
 in this range.  Figure \ref{logybeta} is a set of logarithmic plots which displays the dependence of $D(R)$ on lattice volume, at each value of $\b$.   What we see in these figures is that the data at the different volumes may have converged at $R < 0.6$ fm or so, but at larger separations this is not the case.  At $\b=5.7, 5.8$ the data has dropped below the $x$-axis and gone negative in some region.  At $\b=5.9, 6.0$ the data points remain positive, but there is reason to suspect that at still higher volumes this data might also go negative in the long distance regime. The data is shown on a linear scale
in Fig.\ \ref{linybeta}.     

    It is interesting to see whether the data in the intermediate regime, where we may be seeing scaling, can be fit by
a standard ansatz.  Define the equal-time propagators corresponding to a massive or Gribov form
\bea
           D^{mass}(R,m) &=& \int {d^3 k \over (2\pi)^3} ~ {e^{i\vk \cdot \vx} \over 2 (k^2 + m^2)^\oh } 
\non \\
                                &=& {m \over 4 \pi^2 R} K_1(mR) \ ,
\eea
and
\bea
           D^{Grib}(R,m) &=& \int {d^3 k \over (2\pi)^3} ~ {e^{i\vk \cdot \vx} \over 2 (k^2 + m^4/k^2)^\oh } \ .
\eea
The latter integral can be expressed in terms of MeijerG functions:
\begin{widetext}
\bea
D^{Grib}(r,m) &=& -\frac{m G_{1,5}^{3,1}\left(\frac{m^4 r^4}{256}|
\begin{array}{c}
 \frac{1}{4} \\
 -\frac{1}{4},\frac{1}{4},\frac{3}{4},0,\frac{1}{2} \\
\end{array}
\right)}{8 \sqrt{2} \pi ^2 r}+\frac{3 G_{1,5}^{3,1}\left(\frac{m^4
   r^4}{256}|
\begin{array}{c}
 \frac{3}{4} \\
 \frac{1}{4},\frac{1}{4},\frac{3}{4},0,\frac{1}{2} \\
\end{array}
\right)}{8 \sqrt{2} \pi ^2 m r^3}
-\frac{3 G_{1,5}^{3,1}\left(\frac{m^4
   r^4}{256}|
\begin{array}{c}
 \frac{1}{2} \\
 0,0,\frac{1}{2},\frac{1}{4},\frac{3}{4} \\
\end{array}
\right)}{8 \sqrt{2} \pi ^2 r^2}
\non \\
& &
-3 \left(-\frac{m
   G_{1,5}^{3,1}\left(\frac{m^4 r^4}{256}|
\begin{array}{c}
 \frac{1}{4} \\
 -\frac{1}{4},\frac{1}{4},\frac{3}{4},0,\frac{1}{2} \\
\end{array}
\right)}{8 \sqrt{2} \pi ^2 r}
+\frac{G_{1,5}^{3,1}\left(\frac{m^4
   r^4}{256}|
\begin{array}{c}
 \frac{3}{4} \\
 \frac{1}{4},\frac{1}{4},\frac{3}{4},0,\frac{1}{2} \\
\end{array}
\right)}{8 \sqrt{2} \pi ^2 m r^3}-\frac{G_{1,5}^{3,1}\left(\frac{m^4
   r^4}{256}|
\begin{array}{c}
 \frac{1}{2} \\
 0,0,\frac{1}{2},\frac{1}{4},\frac{3}{4} \\
\end{array}
\right)}{8 \sqrt{2} \pi ^2 r^2}\right) \ .
\eea
\end{widetext}

\begin{figure*}[t!]
\subfigure[~massive]  
{   
 \label{mass}
 \includegraphics[scale=0.6]{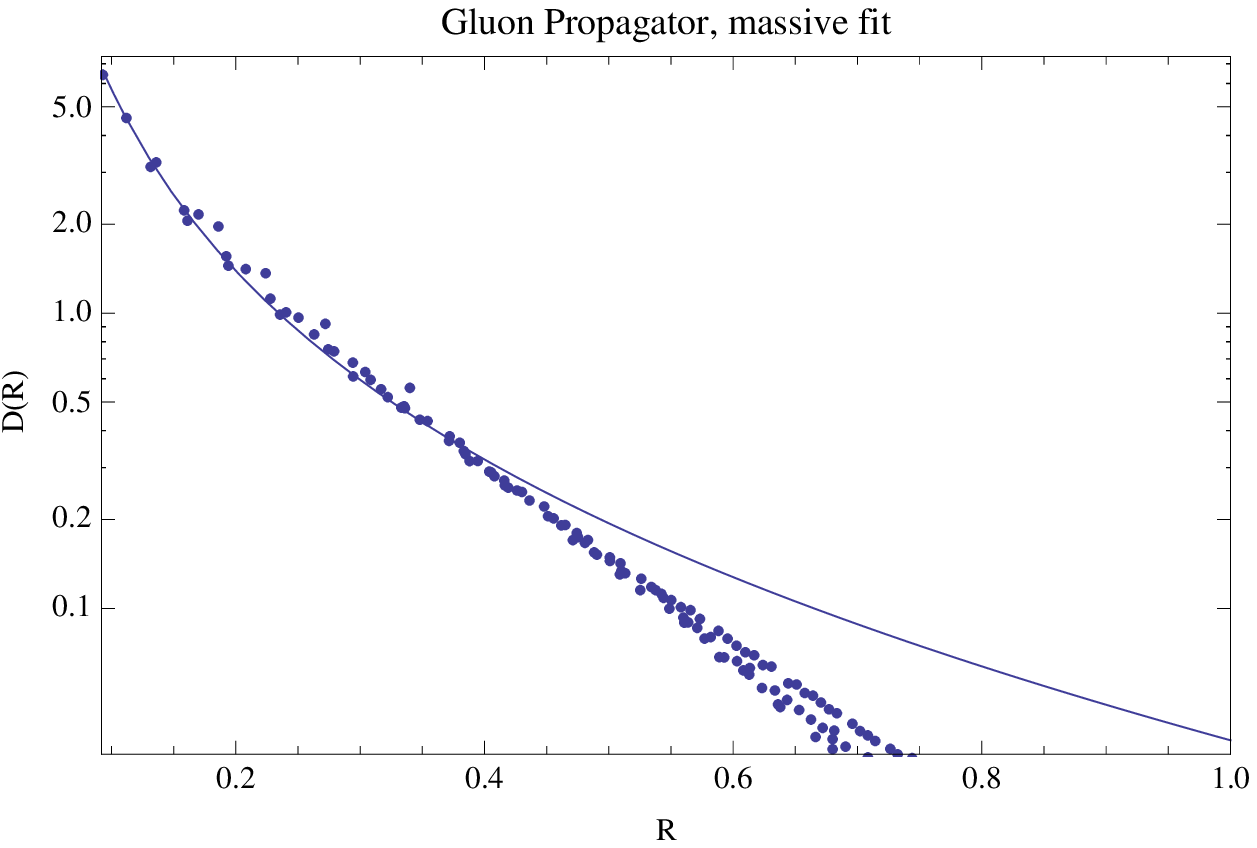}
}
\subfigure[~Gribov]  
{   
 \label{Gribov}
 \includegraphics[scale=0.6]{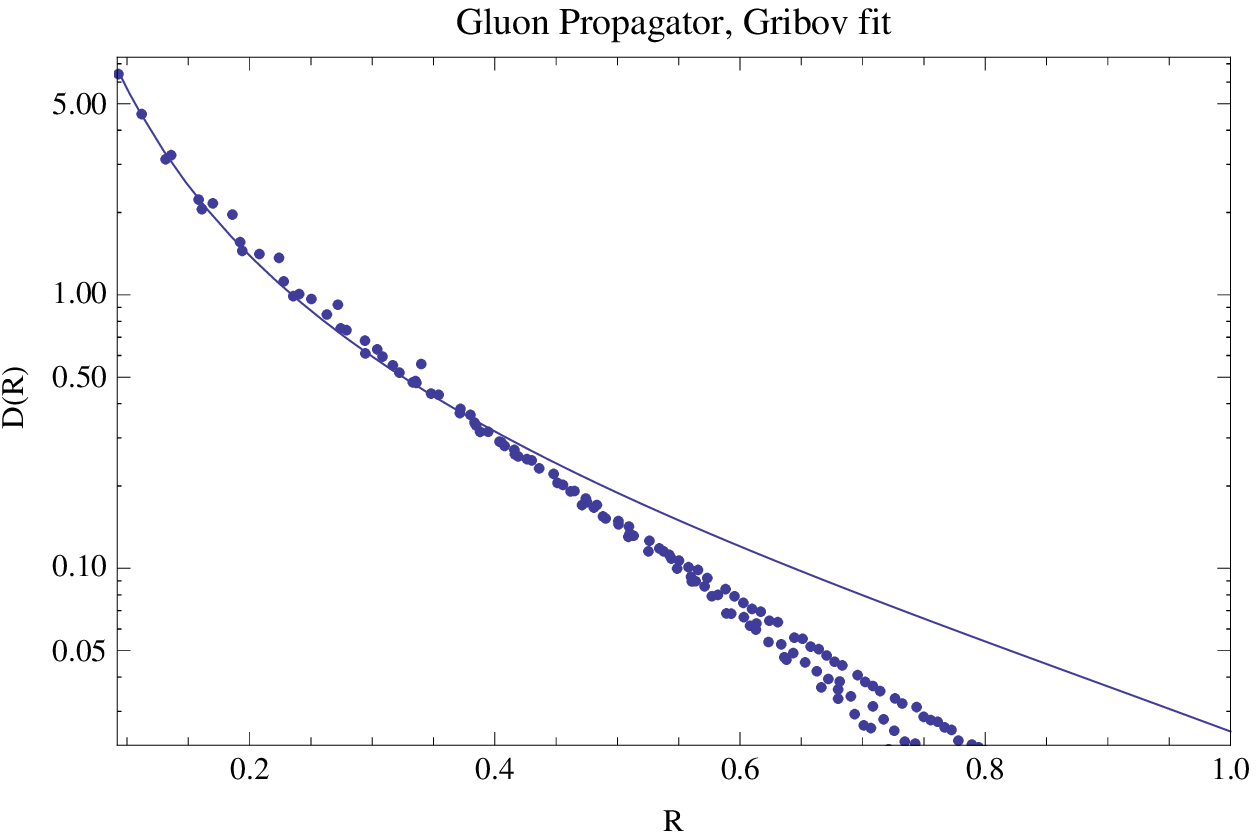}
}
\caption{Comparison of lattice data for the transverse gluon propagator to fits by (a) a propagator with a mass pole;
(b) the Gribov propagator.}
\label{fits}
\end{figure*} 

\begin{figure}[t!]
\subfigure[~ log scale]  
{   
 \label{logy_v30}
 \includegraphics[scale=0.6]{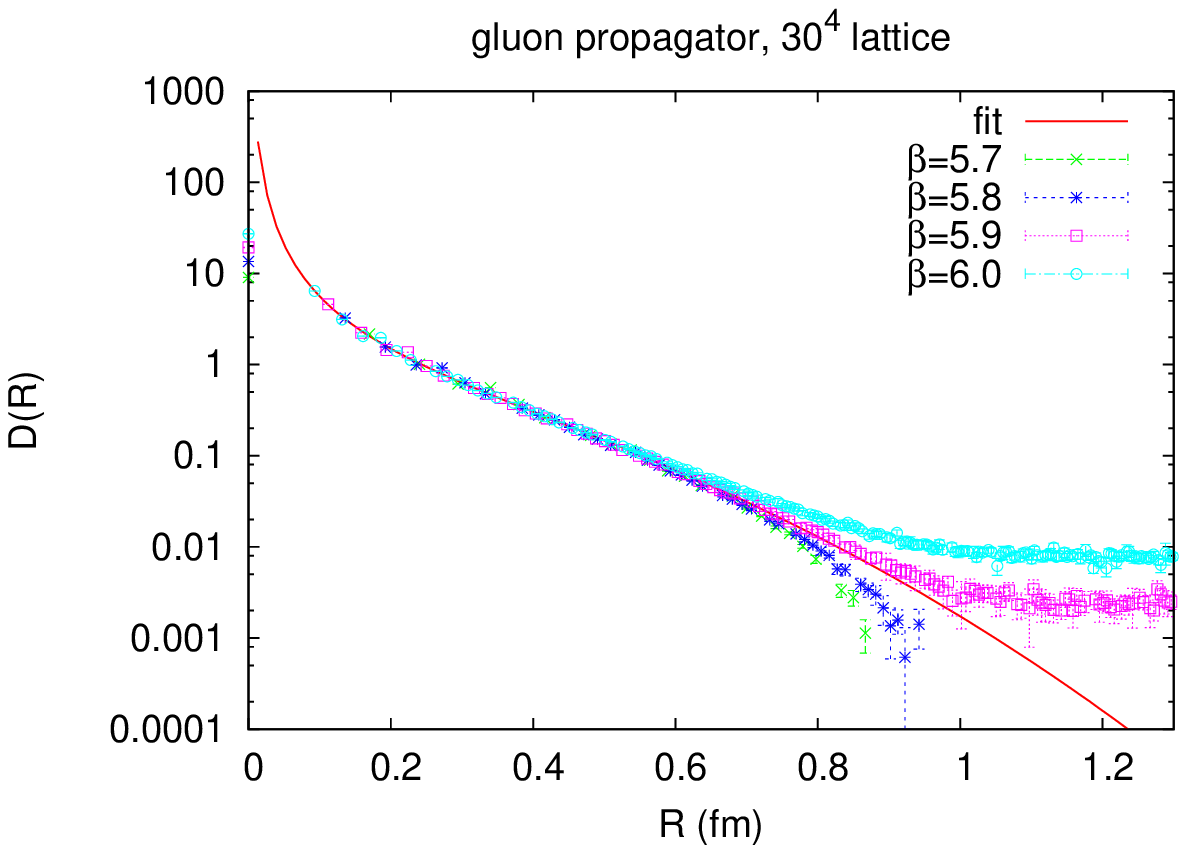}
}
\subfigure[~ linear scale]  
{   
 \label{liny_v30}
 \includegraphics[scale=0.6]{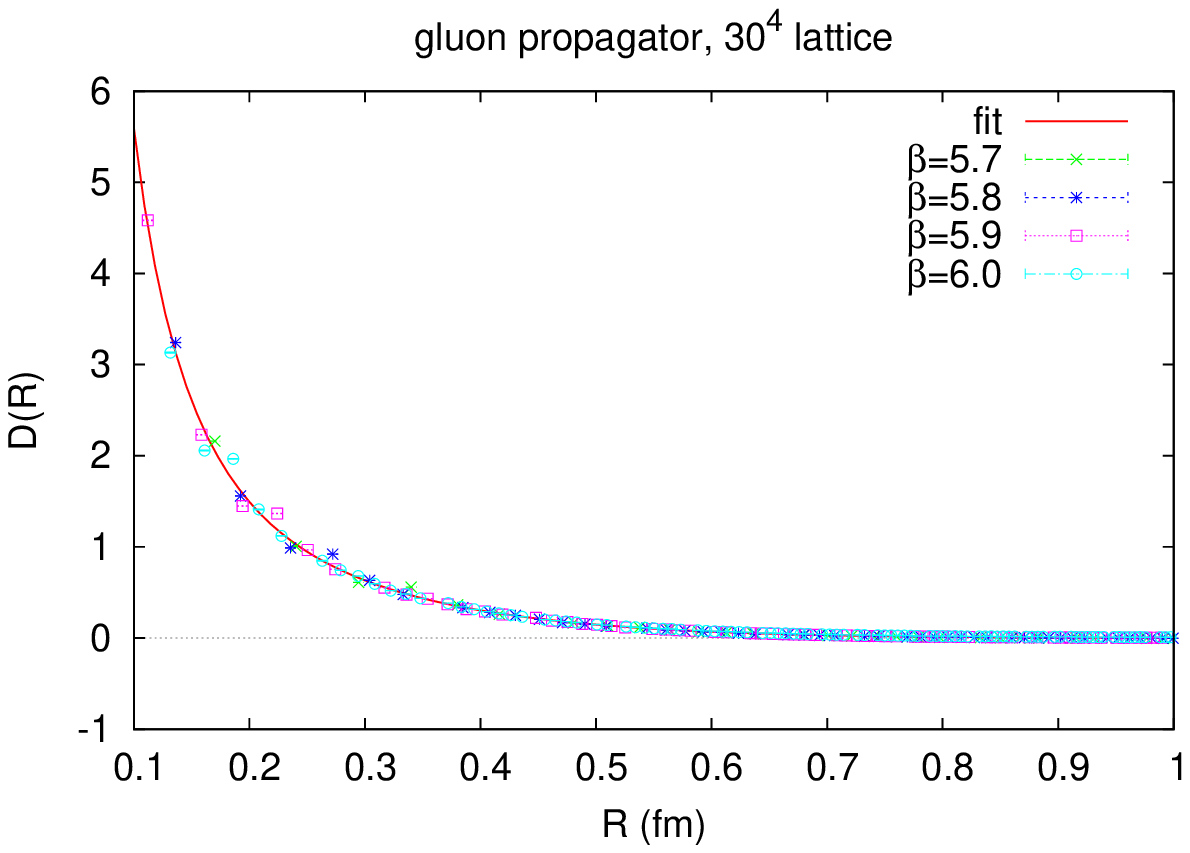}
}
\caption{Comparison of the best fit \rf{trial} to the equal-times transverse propagator data, at various lattice
couplings, on the largest $30^4$ lattice volume.  (a) logarithmic scale; (b) linear scale.}
\label{myfit}
\end{figure}

    Since the data has not convincingly converged at $R>0.6$ fm, we can at least try to fit the data in the range
$0<R \le 0.6$ fm, where we do seem to see both scaling and volume convergence, to either the massive or
Gribov forms, i.e.
\beq
              D(R) = \left\{ \begin{array}{c} 
                                c D^{mass}(R,m) \cr
                                c D^{Grib}(R,m) \end{array} \right. \ ,
\eeq
where $c$ and $m$ are the fitting parameters.  Fig.\ \ref{mass} shows a best fit for the massive propagator, and Fig.\ \ref{Gribov} for the Gribov propagator, on a logarithmic scale.  The data points shown are combined data for all couplings on the largest $30^4$ lattice volume.  Clearly neither of these fits are very convincing.  The fitting functions can get the short distance behavior correctly, where the propagator is large, but then they deviate in the longer range tail, where the propagator is small.  By fitting instead to the logarithm of the data one can get quite a good fit to the tail, but then the fits both go wrong at the short distance end, especially when plotted on a linear scale.  Neither the massive nor the Gribov propagator gives a good account of the data in the full $0<R \le 0.6$ fm interval.  This motivates a search for some other functional form for the propagator. We have found that this form:
\beq
         D^{trial}(R) = c {e^{- b R^2 + a R} \over R^2}
\label{trial}
\eeq 
gives an excellent fit to the data.  In the logarithmic plot in Fig.\ \ref{logy_v30}, and the linear plot
in Fig.\ \ref{liny_v30} we show the data for $D(R)$ at all four $\b$ values at the largest $30^4$ lattice volume.
On the same plots we also display the best fit (solid line) to the data in the interval $0.09 \le R \le 0.6$ fm by $D^{trial}(R)$. The constants which give this fit are
\beq
a = 2.35(23)~{\rm fm}^{-1} ~~,~~ b=5.65(31) ~{\rm fm}^{-2} ~~,~~ c=0.0469(12) \ .
\eeq
There is no particular theoretical justification for the form \rf{trial}.  The ansatz just happens to 
work very well in this particular distance interval, where we have convergent data.  Presumably \rf{trial} is simply a good approximation in this interval to the true, and no doubt very complicated, transverse equal-times gluon propagator, which probably violates positivity at larger distances, and displays logarithmic corrections at shorter distances.  

     It is important to note the striking fact, seen in Figures \ref{logy_v30} and \ref{liny_v30}, that the data at different
values of the lattice spacing, when expressed in physical units, completely overlap up to 0.6 fm or so, beyond which the
data still seems to be volume dependent.   This is not expected.  In lattice simulations we are dealing with the unrenormalized propagator, so data sets at different values of the cutoff ought to differ from one another by multiplicative constants, and in the coupling range shown the largest and smallest lattice spacings differ by almost a factor of two.  Yet there seems to be no trace of
a cutoff-dependent multiplicative constant in the range of lattice couplings we have probed.  The conclusion appears to be that within this range of lattice cutoffs we may have $Z_A \approx 1$, at least in the force renormalization scheme which underlies the Necco-Sommer formula \rf{Necco}, and if that is the case then the bare and renormalized propagators agree in this coupling range.  Of course $Z_A$ must differ very much from unity at sufficiently 
large $\b$ and small lattice spacings.   In any event, the bare gluon propagator for this range of couplings is fit by
\beq
D(R) = 0.0469 {\exp[-5.65 R^2 + 2.35 R] \over R^2} \ .
\label{DR}
\eeq
where $R$ is in fm and $D(R)$ in fm${}^{-2}$.
 
   It should noted that there have been previous studies of both the transverse propagator in Coulomb gauge, carried out in both SU(2) \cite{Langfeld:2004qs,Burgio:2008jr,Burgio:2012bk} and SU(3) \cite{Nakagawa:2011ar}  pure gauge theory.  
These studies computed gluon propagators in momentum space, rather than position space, and the scaling analysis was on time asymmetric lattices.   For these and other technical reasons it is not straightforward to compare our position space results directly with the earlier momentum space studies, and we will not attempt this here. However, given the strong sensitivity of our results to lattice volume, we believe it would be very helpful to carry out further studies of the equal-times transverse gluon propagator in both position and momentum space, on much larger lattices than those used here.

\subsection{\label{ghosts}The ghost propagator}

   We now consider the (bare, unrenormalized) ghost propagator
\bea
         G^{ab}(R) &=& \left\langle \big(\M^{-1}\big)_{a\vx,b\vy} \right\rangle
\non \\
             &=& \d^{ab} {1\over 8}   \left\langle \big(\M^{-1}\big)_{c\vx,c\vy} \right\rangle \ ,
\eea   
where the Faddeev-Popov operator on any given timeslice of the lattice is
\bea
\M_{a\vx, b\vy} &=& \mbox{ReTr}\sum_{k=1}^3 \Big[ \{t^a,t^b\}(U_k(\vx)+U_k(\vx-\hat{k})) \d_{\vx \vy} 
\non \\ 
& & - 2t^b t^a U_k(\vx) \d_{\vx+\hat{k},\vy} - 2t^a t^b U_k(\vx-\hat{k}) \d_{\vx-\hat{k},\vy} \Big] \ .
\non \\
\eea   
with the $U_k(\vx)$ being the spacelike link variables on that timeslice.
However, $\M$ has eight eigenvectors with zero eigenvalues ($c=1-8$)
\beq
            \psi^{(c)}_{a\vx} = {1\over L^{3/2}} \d_{ac} ~~~\mbox{with}~~~  \M_{a\vx,b\vy} \psi^{(c)}_{b \vy} = 0 
\eeq
associated with a remnant global SU(3) color symmetry, and is therefore not invertible as it stands. We must therefore
invert $\M$ on a subspace orthogonal to these zero modes.  An alternate procedure is to invert the Faddeev-Popov
operator in momentum space, as in \cite{Burgio:2008jr,Burgio:2012bk,Nakagawa:2009zf}.  Nakagawa et al.\ \cite{Nakagawa:2009zf} find in momentum space, for the SU(3) group
\beq
           G(k) \propto {1 \over k^{2.44}}  ~~~ \mbox{(infrared)}
\eeq
Burgio et al.\ \cite{Burgio:2008jr,Burgio:2012bk} and Langfeld and Moyaerts \cite{Langfeld:2004qs} find nearly the same power dependence in SU(2).  Translating
this into position space, we write
\beq
           G_c(R) \equiv gG(R) = {\sqrt{6} \over 8} {c(\b) \over R^{0.56}} \label{Gc} 
\eeq
The motivation for multiplying the ghost propagator by coupling $g$ is that, as we have noted above, the
combination $g G(R) Z_A^{-1}$, where $gG$ is the product of the bare lattice coupling and ghost propagator, is equal to the product of the renormalized coupling and ghost propagator.  We have already
seen evidence that $Z_A \approx 1$ in the range of couplings $\b \in 5.7-6.0$.   We therefore treat $c(\b) \approx c$ as a constant in this coupling range, to be fixed by fitting to some physical quantity.  As mentioned below \rf{KAA}, this constant may also absorb a contribution correcting for the factorization of the operator expectation values shown in Fig.\ \ref{decomp}.

\section{\label{st} The Static Quark Potential}

   We now define the matrix $[H]$ to have elements $H_{mn} = \langle m|H|n \rangle$ where $H_{00}$ and
$H_{n0}=H_{0n}^*$ were defined in eqs.\ \rf{H00} and \rf{H10} respectively, and 
\bea
H_{nm} &=& H_{nm}^{kin} + H_{nm}^{coul} + H_{nm}^{np} + H_{nm}^{KGG}   \ ,
\eea
where expressions for $H_{nm}^{kin}, H_{nm}^{coul}, H_{nm}^{np}, H_{nm}^{KGG}$ have been given
in eqs.  \rf{H11kin}, \rf{H11coul}, \rf{H11np}, \rf{H11KGG}.  All of these matrix elements, as well as the elements
$O_{mn}$ of the overlap matrix $[O]$ defined in \rf{Omn} depend on the parameter $a$ that appears in the
one-gluon wavefunctions $\Psi$, as well as the parameter $c$ in the ghost propagator.

\begin{figure}[t!]
\centerline{\scalebox{0.6}{\includegraphics{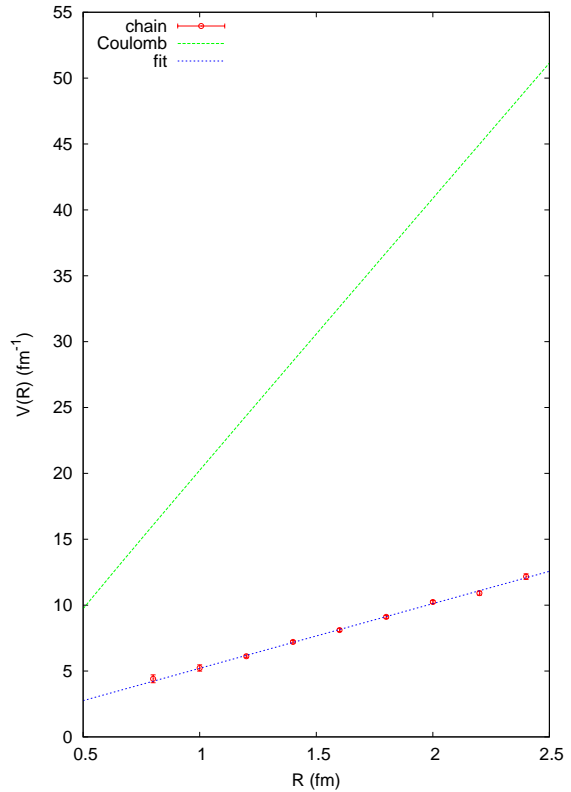}}}
\caption{The Coulomb and gluon chain potentials.}
\label{potential}
\end{figure} 

   The procedure, for any given ghost parameter $c$ and quark-antiquark separation $R$, is to compute the energy spectrum of the Hamiltonian in a truncated basis
which spans the Hilbert subspace containing the non-orthogonal zero and one-gluon states $\{ |n\rangle \}$, and vary
the parameter $a$ to minimize the lowest energy eigenvalue.  As we have already mentioned, a possible procedure
is to construct an orthonormal basis from the $\{ |n\rangle \}$ states via Gram-Schmidt orthogonalization, compute
the Hamiltonian matrix elements in this basis, and then diagonalize the Hamiltonian.  An equivalent method is
to solve the generalized Hamiltonian equation $[H] \vec{v}_n = E_n [O] \vec{v}_n$; this gives the same spectrum
of energy eigenvalues.  This procedure is carried out for a range of quark-antiquark separations $R$, and parameters
$c$.  If the lowest energy $E_{min}(R)$ is linear as a function of $R$, then we choose $c$ to get $\s = E_{min}(R)/R$
equal to the asymptotic string tension. In our calculation we have chosen indices $\a,\b$ in the wavefunction (see \rf{Fnm}) in the range $\a=1,2,3,4, ~ \b=1,2,3$, so counting the zero-gluon state this is thirteen states in all.

    Our final result, and the main result of this paper, is shown in Fig.\ \ref{potential}.  Here we plot the variational/truncated basis estimate for the static quark potential $V(R)$ up to $R=2.4$ fm, in intervals of 0.2 fm, as compared to the Coulomb potential of eq.\  \rf{cpot}.  We have found that as we increase the number of states in the truncated basis from two to thirteen, the ground state energy converges at around six or seven states.  Error bars, comparable to symbol size, are estimates of the error deriving from numerical integrations.    The solid line through the computed points is a best fit by the function
\beq
            V^{fit}(R) = \s^{fit} R +  \d \ .
\eeq
and we find
\bea
           \s^{fit} &=& 4.91(9) ~ \text{fm}^{-2}  
\non \\
                      &=& (437(4) ~ \text{MeV})^2 \ .
\eea
which can be compared to the accepted value of  $\s= (440 ~ \text{MeV})^2$.  Of course, the parameter in the ghost
propagator $c=3.5$ was chosen to obtain this result, but what is non-trivial is the fact that the potential remains linear, despite deviating very strongly from the non-perturbative Coulomb potential, which we recall had a string tension four times greater than the asymptotic string tension.

\section{\label{conclude}Conclusions}
 
     We have shown how the inclusion of a single constituent gluon in the static quark-antiquark state can preserve the
linear dependence of the energy on quark separation $R$, at least up to $R=2.4$ fm, while drastically reducing the string tension from the
pure Coulomb value of $\s_{coul}= (891~\mbox{MeV})^2$.  To show that the reduction is by precisely the right amount,
namely a factor of four, will require further numerical studies of the ghost and transverse gluon propagators.

     The next step in our program is to apply our variational approach to gluelumps, glueballs, and heavy (but not static)
quark-antiquark bound states.  For quarkonium states the diagrams are essentially the same as those
shown in section \ref{matrix}, only replacing the static quark lines with dynamical quarks and antiquarks, and of course there will be wavefunctions for relative positions and spin of quark, antiquark, and (zero or one) constituent gluons.  Likewise, for the glueball states, the static quark lines are replaced by gluon propagators.  For these states, where no constituent is located at a fixed position, it is preferable to carry out the calculation of Hamiltonian matrix elements in momentum rather than position space.  This takes advantage of overall momentum conservation, and greatly reduces the dimensionality of integrals that must be carried out numerically.  We hope to report on work along these lines
in a future publication.  \\

\acknowledgments{This work was supported in part by the U.S. Department of Energy under Grants No. 
DE-FG03-92ER40711 (JG) and  DE-FG0287ER40365 (APS). The work was authored in part by Jefferson Science 
Associates, LLC under U.S. DOE Contract No. DE-AC05-06OR23177.}

\appendix*

\section{Comparison with Dyson-Schwinger equations}

In this appendix we compare our results for the ghost and Coulomb propagators derived from lattice simulations with the    predictions of Dyson-Schwinger (DS) equations~\cite{Szczepaniak:2001rg,Feuchter:2004mk,Schleifenbaum:2006bq, Epple:2007ut}. 
 In the one-loop, rainbow-ladder approximation,   the renormalized DS equation for the momentum space ghost dressing function  $d(p) \equiv  gG(p) p^2$ is given by 
\begin{equation} 
\frac{1}{d(p)} = \frac{1}{d(\mu)} -  I_d[d(k)] + I_d[d(\mu)] \label{d} \ ,
\end{equation} 
where 
\begin{equation} 
I_d[d(k)] = N_C  \int \frac{d^3k}{(2\pi)^3} \frac{1 - (\hat p\cdot \hat k)^2}{2\omega(k-p)} \frac{d(k)}{(p-k)^2}, 
\end{equation}
and $2\omega(p)= D^{-1}(p)$ is the inverse momentum space gluon propagator. Given the function $\omega(p)$, the solution of  (\ref{d}) depends on the value of  $d(\mu)$ at an arbitrary renormalization point, which we choose as $\mu = 1 ~\mbox{fm}^{-1}$.  
Specifically, the  behavior of the ghost dressing function in the IR, i.e.\ for $p\to 0$, depends on the value of $d(\mu)$  and the IR behavior of  $\omega(p)$. If $\omega(0)$ is finite then so is $d(0)$, providing that $d(\mu)$ is less then some critical value $d_c$ which depends on $\omega$. As $d(\mu)$ increases towards the critical value the  ghost dressing function becomes IR enhanced, and $d(0)$ becomes 
 infinite for $d(\mu) =   d_c$ with  $d(p) \propto 1/p^{1/2}$ as  $p\to 0$. 
   In general,  if $\omega(p)$ behaves in the IR as 
   \begin{equation} 
   \omega(p) \propto 1/p^\beta \label{iro} \ ,
   \end{equation}
the ghost dressing function  behaves as 
   \begin{equation} 
   d(p) \propto   1/p^{(1 + \beta)/2} \label{ird} \ ,
   \end{equation} 
so long as $d(\mu)$ is equal to the critical value. The  large-$R$ behavior of the ghost propagator $gG(R)$ in position space, which is implied by (\ref{ird}),  is 
 \begin{equation} 
 gG(R) \propto 1/R^{1/2 - \beta/2} \ . \label{girr} 
 \end{equation} 
   For any $\beta$ and $d(\mu) < d_c$,  $d(0)$ is finite.  However, there is a range of low momenta where $d(p)$ follows the power-law behavior given by (\ref{ird}). The lower limit in this range approaches 
     $p=0$ as $d(\mu) \to d_c$. Similarly, (\ref{iro}) with $\beta > 0$  implies that in position space at large  distances ($R\to  \infty$)  the gluon propagator behaves as 
   \begin{equation} 
D(R) \propto 1/R^{ 3 + \beta } \ .
\end{equation} 
We  note that if $D(R) > 0$  at all distances, $\omega(0)$ is finite, while an IR enhanced $\omega$ with   $\beta  > 0$ implies that the gluon propagator is positivity violating.  
The momentum space Coulomb potential is proportional to the square of the ghost propagator   and the Coulomb form factor $f$, 
\begin{equation} 
C_F g^2 K(p) = - \frac{d^2(p) f(p)}{p^2} \ ,  
\label{vSD} 
\end{equation} 
which satisfies the following DS equation, 
\begin{equation} 
f(p) = f(\mu) + I_f[f(p)] - I_f[f(\mu)] \label{fSD} \ ,
\end{equation} 
with 
\begin{equation} 
I_f[f(k)] = N_C  \int \frac{d^3k}{(2\pi)^3} \frac{1 - (\hat p\cdot \hat k)^2}{2\omega(k-p)} \frac{d^2(k)f(k)}{(p-k)^2} \ . 
\end{equation}
It follows from (\ref{fSD}) that $f(p)$ is determined up to an overall normalization.


  A good approximation to the momentum space lattice propagator 
   can be obtained using a simple form motivated by the Gribov ansatz  
      \begin{equation}
\omega(p) = \sqrt{(p/a)^2 + m_g^2 (m_g/p)^{2\beta} } \label{gluonSD}  \ .
\end{equation} 
 The functional form adopted in \rf{DR}, i.e.\ a positive $D(R)$, implies   $\beta=0$.  In this case the DS equation (\ref{d}) results in a ghost propagator which at large $R$ behaves as $gG(R) = (\sqrt{6}/8) A R^{-1/2}$  with $A \approx 1.5$. 
 The power behavior is close to the lattice fit (\ref{Gc}), but the magnitude 
   is approximately a factor of two smaller than what is required (according to the analysis in this article) to fit the asymptotic string tension. It follows from (\ref{d}) that to increase the value of the ghost propagator  it is necessary to enhance  
 $\omega$ in the IR, and this in turn implies positivity violation. 
  The lattice data on the gluon propagator does indeed indicate positivity violation.  While the analytical form in  \rf{DR} is  acceptable in the analysis of the gluon chain spectrum, comparison with the DS equation requires taking positivity violation into account.

\bibliography{chain}

\begin{thebibliography}{10}

\bibitem{Greensite:2001nx}
J.~Greensite and C.~B. Thorn,
\newblock JHEP {\bf 0202}, 014 (2002), arXiv:hep-ph/0112326.

\bibitem{Greensite:2014bua}
J.~Greensite and A.~P. Szczepaniak,
\newblock Phys.Rev. {\bf D91}, 034503 (2015), arXiv:1410.3525.

\bibitem{Nakagawa:2006fk}
Y.~Nakagawa, A.~Nakamura, T.~Saito, H.~Toki, and D.~Zwanziger,
\newblock Phys.Rev. {\bf D73}, 094504 (2006), arXiv:hep-lat/0603010.

\bibitem{Voigt:2008rr}
A.~Voigt, E.-M. Ilgenfritz, M.~Muller-Preussker, and A.~Sternbeck,
\newblock Phys.Rev. {\bf D78}, 014501 (2008), arXiv:0803.2307.

\bibitem{Burgio:2012bk}
G.~Burgio, M.~Quandt, and H.~Reinhardt,
\newblock Phys.Rev. {\bf D86}, 045029 (2012), arXiv:1205.5674.

\bibitem{Greensite:2004ke}
J.~Greensite, S.~Olejnik, and D.~Zwanziger,
\newblock Phys.Rev. {\bf D69}, 074506 (2004), arXiv:hep-lat/0401003.

\bibitem{Greensite:2003xf}
J.~Greensite and S.~Olejnik,
\newblock Phys.Rev. {\bf D67}, 094503 (2003), arXiv:hep-lat/0302018.

\bibitem{Zwanziger:2002sh}
D.~Zwanziger,
\newblock Phys.Rev.Lett. {\bf 90}, 102001 (2003), arXiv:hep-lat/0209105.

\bibitem{Greensite:2009mi}
J.~Greensite and S.~Olejnik,
\newblock Phys.Rev. {\bf D79}, 114501 (2009), arXiv:0901.0199.

\bibitem{Szczepaniak:2005xi}
A.~P. Szczepaniak and P.~Krupinski,
\newblock Phys.Rev. {\bf D73}, 034022 (2006), arXiv:hep-ph/0511083.

\bibitem{Burgio:2008jr}
G.~Burgio, M.~Quandt, and H.~Reinhardt,
\newblock Phys.Rev.Lett. {\bf 102}, 032002 (2009), arXiv:0807.3291.

\bibitem{Nakagawa:2009zf}
Y.~Nakagawa {\em et~al.},
\newblock Phys.Rev. {\bf D79}, 114504 (2009), arXiv:0902.4321.

\bibitem{Zwanziger:1998ez}
D.~Zwanziger,
\newblock Nucl. Phys. {\bf B518}, 237 (1998).

\bibitem{Feinberg:1977rc}
F.~L. Feinberg,
\newblock Phys.Rev. {\bf D17}, 2659 (1978).

\bibitem{Cucchieri:2000hv}
A.~Cucchieri and D.~Zwanziger,
\newblock Phys.Rev. {\bf D65}, 014002 (2001), arXiv:hep-th/0008248.

\bibitem{Watson:2007mz}
P.~Watson and H.~Reinhardt,
\newblock Phys.Rev. {\bf D76}, 125016 (2007), arXiv:0709.0140.

\bibitem{Davies:1987vs}
C.~Davies {\em et~al.},
\newblock Phys.Rev. {\bf D37}, 1581 (1988).

\bibitem{Necco:2001xg}
S.~Necco and R.~Sommer,
\newblock Nucl.Phys. {\bf B622}, 328 (2002), arXiv:hep-lat/0108008.

\bibitem{Langfeld:2004qs}
K.~Langfeld and L.~Moyaerts,
\newblock Phys. Rev. {\bf D70}, 074507 (2004), arXiv:hep-lat/0406024.

\bibitem{Nakagawa:2011ar}
Y.~Nakagawa, A.~Nakamura, T.~Saito, and H.~Toki,
\newblock Phys. Rev. {\bf D83}, 114503 (2011), arXiv:1105.6185.

\bibitem{Szczepaniak:2001rg}
A.~P. Szczepaniak and E.~S. Swanson,
\newblock Phys.Rev. {\bf D65}, 025012 (2002), arXiv:hep-ph/0107078.

\bibitem{Feuchter:2004mk}
C.~Feuchter and H.~Reinhardt,
\newblock Phys.Rev. {\bf D70}, 105021 (2004), arXiv:hep-th/0408236.

\bibitem{Schleifenbaum:2006bq}
W.~Schleifenbaum, M.~Leder, and H.~Reinhardt,
\newblock Phys.Rev. {\bf D73}, 125019 (2006), arXiv:hep-th/0605115.

\bibitem{Epple:2007ut}
D.~Epple, H.~Reinhardt, W.~Schleifenbaum, and A.~Szczepaniak,
\newblock Phys.Rev. {\bf D77}, 085007 (2008), arXiv:0712.3694.

\end{thebibliography}

\end{document}